\documentstyle[12pt] {article}

\input epsf

\begin{document}\setlength{\unitlength}{1mm}
\def\question#1{{{\marginpar{\small \sc #1}}}}
\newcommand{\QCD}{{ \rm QCD}^{\prime}}
\newcommand{\MSSM}{{ \rm MSSM}^{\prime}}
\newcommand{\eq}{\begin{equation}}
\newcommand{\en}{\end{equation}}
\newcommand{\bino}{\tilde{b}}
\newcommand{\tsquark}{\tilde{t}}
\newcommand{\gluino}{\tilde{g}}
\newcommand{\photino}{\tilde{\gamma}}
\newcommand{\wino}{\tilde{w}}
\newcommand{\mtilde}{\tilde{m}}
\newcommand{\higgsino}{\tilde{h}}
\newcommand{\gsi}{\,\raisebox{-0.13cm}{$\stackrel{\textstyle>}
{\textstyle\sim}$}\,}
\newcommand{\lsi}{\,\raisebox{-0.13cm}{$\stackrel{\textstyle<}
{\textstyle\sim}$}\,}
\rightline{CERN-TH/98-196}
\rightline{hep-ph/9806519}
\rightline{June, 1998}
\baselineskip=18pt
\vskip 0.6in
\begin{center}
{ \LARGE The Two-Loop Finite-Temperature Effective Potential of the MSSM and Baryogenesis}\\
\vspace*{0.6in}
{\large Marta Losada\footnote{On leave of absence
from the Universidad Antonio Nari\~{n}o, Santa Fe de
Bogot\'a, COLOMBIA.}} \\
\vspace{.1in}
{\it CERN Theory Division \\ CH-1211
Geneva 23, Switzerland}\\
\vspace{.2in}
\vspace{.1in}
\end{center}
\vspace*{0.05in}
\vskip  0.2in  

Abstract: We construct an effective three dimensional theory for
the MSSM at high temperatures  in the
limit of large-$m_{A}$.
 We analyse the two-loop effective potential of
the 3D theory for the case of a light right handed stop to determine
the precise region in the $m_{h}$-$m_{\tilde{t}_{R}}$ plane
for which the sphaleron constraint  for preservation
of the baryon asymmetry is satisfied. We also compare
with results previously obtained using 3D and 4D  calculations
of the effective potential. A two-stage phase transition still persists
for a small range of values of $m_{\tilde{t}_{R}}$. The allowed region requires a
value of $m_{\tilde{t}_{R}}\lsi m_{t}$ and $m_{h} \lsi 100$ $(110)$ GeV for $m_{Q}=300$ GeV ($1$ TeV).

\thispagestyle{empty}
\newpage
\addtocounter{page}{-1}
\newpage

\section{Introduction}

The analysis of the electroweak phase transition in the MSSM
has been the subject of intensive study in recent years. The main motivation
is the generation (and preservation) of a possible baryon asymmetry at the
electroweak scale \cite{kuzmin, shaposhnikov} (for reviews, see \cite{ckn}-\cite{quiros}).
Many different contributions have given a clearer
idea as to where in parameter space  the condition of a strong enough
 first-order
phase transition can occur   in order for electroweak
baryogenesis to be possible. Mainly two different analytic approaches have been
used in the analysis of the phase transition for the MSSM. The first one
involves the evaluation of the effective potential in the 4D theory.
The one-loop analysis \cite{myint}-\cite{zwirner2} led to the conclusion that
low values of the ratio of the vacuum expectation values of the two Higgs doublets $\tan\beta = {v_{2}\over v_{1}}$, and large values of the pseudoscalar mass $m_{A}$ were favoured.
More recently the effects of a very light right handed stop with a soft SUSY-breaking mass value of $m_{U}^{2} \lsi 0$
were shown to strongly  strengthen  the phase transition by
enhancing the cubic term in the effective potential \cite{Carena1, delepine}.
As first pointed out by Espinosa \cite{Espinosa}, two-loop corrections were also
 shown to be very important in order to
increment the allowed parameter space for which  electroweak baryogenesis
can take place.  In particular 
 two-loop QCD corrections from stops can strongly affect the
value of the scalar field at the phase transition \cite{Espinosa}-\cite{Laine2}.
A precise determination of the region in parameter space for which electroweak
baryogenesis is viable was
 done by Carena et al. \cite{Carena2}, using a two-loop calculation
in 4D, with a light right handed stop, and a heavy third generation left squark doublet, 
which is decoupled from the thermal bath. They conclude that
the Higgs mass must be lower than $105$ GeV and that
the right stop mass must be in the range of $110$-$160$ GeV if 
absolute stability of the physical vacuum is required.
A very recent paper by Cline and Moore \cite{cm} use the two-loop effective
potential in 4D, fully incorporating squark and Higgs boson mixing,
to determine the allowed region in parameter space. Their results are in good 
agreement with those of ref. \cite{Carena2}.

In  the second approach for the analysis of the phase transition,
 the aim is to separate
the perturbative and non-perturbative aspects of the theory.
   The purely perturbative component of the calculation is performed by constructing an effective
3D theory for the model under consideration \cite{ginsparg}-\cite{Nieto}.  The  parameters
in the 3D Lagrangian
are obtained using dimensional reduction at high temperature
by matching the static Green's functions in the two theories,
to a given order in the perturbative expansion, by  integrating out the non-zero Matsubara modes with
masses of the order of $\pi T$, where $T$ is the temperature. A further  reduction can also be performed noting that
 some of the static modes  in the theory have acquired thermal masses proportional to a gauge coupling multiplied by the temperature, $\sim g_{w}T, g_{s}T$.
These so-called heavy particles can then be integrated out as well.
The effective potential calculated in the 3D theory reproduces the results
obtained with the 4D effective potential. References \cite{KLRS}, \cite{Laine}-\cite{Losada1} give more details concerning the construction of
effective theories for both  the Standard Model and the MSSM. This approach
simplifies the theory as it is now purely bosonic, which facilitates
lattice simulations. Numerical studies of the
reduced theory will take into account the non-perturbative
effects.
We refer the reader to the above publications for further discussion
regarding strengths and
weaknesses of 4D and 3D calculations. 

For the generic case in which there is a single light scalar doublet field at the 
 phase transition the constraint on strength of the transition is translated into an upper bound on
the ratio of the 3D Higgs self-coupling and the square 3D gauge coupling ${\lambda_{3}\over g_{w_{3}}^{2}}$ \cite{KLRS}. This ratio has a weak dependence on the temperature for values close to the critical temperature of the phase transition
for the allowed range of values of the Higgs mass \cite{Losada2}.
In addition, the 3D scalar and gauge couplings are renormalization group invariant.
 This implies that a 1-loop matching of the 3D coupling constants and masses
to the physical parameters and the temperature
suffices to determine the strength of the phase transition, using the
constraint given by the non-perturbative analysis of the phase
transition for a single light scalar field \cite{KLRS}.

To clarify an essential point, we recall that the critical temperature for
the transition from the high temperature minimum to the
standard electroweak minimum  is
obtained from the condition that the value of the effective potential
at these two minima are equal, $V(0,T) = V(\phi_{c},T)$. Therefore we insist that
 the value of the critical temperature does depend on a
precise determination of the 3D mass parameter, which is not
renormalization group invariant. The scalar mass requires ultraviolet renormalization and a 2-loop calculation (in 4D) must be performed even for the case of a single light scalar at the phase
transition. However, since the ratio of the vacuum expectation value of the scalar field to the temperature, ${\phi_{c}\over T_{c}}$ (or equivalently, ${\lambda_{3}\over g_{w_{3}}^{2}}$), which determines
the rate of sphaleron transitions in the broken phase  has
only a weak dependence, in this case, on the temperature, the two-loop calculation is not necessary.

In the initial analysis for the MSSM using the 3D approach \cite{Laine}-\cite{Losada2}, the light stop
scenario could only be investigated for values of the right handed stop soft SUSY-breaking mass of $m_{U} \gsi 50$ GeV. For lower
values of the stop mass the perturbative procedure of integrating out
the ``heavy" modes starts to break down as the relevant expansion parameter is of the form ${g_{s_{3}}^{2}\over m_{U_{3}}}$. In addition, the numerical
constraint from non-perturbative studies is no longer valid and new studies
 that include the effect of the SU(3) gauge fields must be performed.

An  effective Lagrangian for a
light Higgs and a light stop  must be constructed  to analyse the phase transition for lower values of $m_{U}$
within the 3D approach \cite{Laine2}. A surprising result of the perturbative analysis in this scenario
was that a possible two-stage phase transition could take place, in which
the Universe would first undergo a transition to a colour-breaking vacuum  and, at a lower
temperature, another transition to the physical vacuum would occur.
In fact, the work of Bodeker et al. was the first
to point out that, for  a small range
of values of  $m_{\tilde{t}_{R}}$,
the phase transition could occur in two stages for the MSSM.
This analysis was based on a 2-loop calculation
of the effective potential in 3D including the leading corrections
in the dimensional-reduction procedure.

It is of course of great interest to exactly identify  the range of parameter
space for which this two-stage phase transition can occur. In this case
it is necessary to determine the critical temperatures very precisely
for the two possible directions of the transition. The most
 relevant quantities that determine the critical temperatures
are the 3D mass parameters for the Higgs doublet and the right handed stop. These masses
depend logarithmically on two parameters,  $\Lambda_{H_{3}}$ or $\Lambda_{U_{3}}$,
which can only be determined exactly  from the 4D theory. In the initial
reduction implemented by \cite{Laine2}, the exact values of  $\Lambda_{H_{3}}$ and $\Lambda_{U_{3}}$
were not determined. An estimate was used instead, based on the
value of the corresponding parameter in the Standard Model.
Here we employ a combination of 4D and 3D calculations of the
effective potential to obtain the exact values of  $\Lambda_{H_{3}}$ and $\Lambda_{U_{3}}$.

Undoubtably, many of the questions that arise from
the limitations of perturbation theory will
only be answered when  the corresponding lattice
calculations are done. However, the first non-perturbative
results  show that the 2-loop results for the
strength of the phase transition are conservative
in the bounds they impose on the allowed masses for
the Higgs and the light stop \cite{Laine2}. These
results were obtained using  a simplified model in the reduction procedure.
For more complicated
initial Lagrangians  the final effective 3D theory is characterized by the
same couplings and masses. We would like to point
out a few of the features that can
be affected in the perturbative calculation
of the effective theory and their consequent effect on baryogenesis.
Additionally, in order to apply the non-perturbative results, a precise mapping
of the 4D to 3D parameters is needed, which is presented in this paper. The results obtained with the perturbative 2-loop effective potential
presented here
can then be compared with the lattice analysis.

The procedure of constructing an effective 3D theory is based on the 
mass hierarchy which appears at finite temperature. The validity of the results
depends on an adequate expansion parameter and the suppression of the  
higher order terms in the 3D theory. The value of the masses of the particles which
are integrated out will define the regime of validity of the approach.
As mentioned above, previous analyses \cite{Laine}-\cite{Losada2} of the 3D
theory integrating out the right handed stop claimed that
for $m_{U} \gsi 50$ GeV  the higher order terms were suppressed and consequently the effective theory was an
adequate description. We compare the results obtained
for the ratio of the vacuum expectation value of the Higgs field to the temperature using the two-loop effective potential with and without integrating
out the right handed stop to determine more precisely the value
of the right handed stop mass for which the effective theory is no longer valid. We find that for values of $m_{U} \lsi 123$ GeV the results
obtained having integrated out the right handed stop are unreliable.

The paper is organized as follows: in section 2.1 we present the dimensional
reduction to the effective bosonic theory at one-loop. Section 2.2 presents
a further one-loop reduction in the 3D theory,  eliminating the heavy fields. In section 2.3 we give the expression for the 2-loop
unresummed effective potential in 4D, which is necessary for evaluating
 $\Lambda_{H_{3}}$ and $\Lambda_{U_{3}}$. The contribution to the 3D effective
potential from the ``heavy" particles that were
integrated out at the second stage is given in section 2.4. Section 2.5 discusses
the relevant zero-temperature effects that  must be included in 
our analysis.
In section 3 we present our results for the
critical temperatures and the strength of the phase transition. The
allowed region for electroweak baryogenesis to occur is also given here. A comparison of
the results obtained from the effective potential with and without integrating out
the right handed stop is given in this section. Finally, in section 4, we conclude.
The appendix  contains the relevant formulae for the case in which the
right handed stop is integrated out.

\section{Dimensional Reduction}

We will now perform dimensional reduction by matching, as has been previously
 done in refs. \cite{KLRS},\cite{Laine}-\cite{Losada1}  for different models. Our initial 4D Lagrangian
corresponds to the MSSM in the large-$m_{A}$ limit.
 The particles that
contribute to the thermal bath are the Standard Model particles plus 
third-generation squarks: $\tilde{t}_{L}, \tilde{b}_{L}, \tilde{t}_{R},
 \tilde{b}_{R}$. We will only consider here the case
of zero squark mixing. The results for non-zero squark mixing will
be presented elsewhere\footnote{ The effects of non-zero squark mixing in the
case of a relatively light left squark doublet can complicate
the calculation considerably. For a heavy left squark doublet at large values of 
the mixing parameters, the two-stage scenario is not realized \cite{Carena2}. However, this effect
may not persist for lower values of the mass of the left doublet as
the contribution to the thermal mass of the scalars is
changed, see \cite{Laine, Losada1,  Losada4}. On the other hand, as was noticed in previous studies of the phase transition, a non-zero value of the mixing parameters
 always
weakens the strength of the transition. So lower
values of the Higgs mass  or the right handed stop mass are
necessary to enhance the strength of the transition for
non-zero squark mixing. Thus our results give
upper 
bounds on the scalar masses. } \cite{Losada4}. There are two stages of reduction. The first one corresponds
to the integration out of all non-zero Matsubara modes, that is
with a thermal mass of the order of $\sim \pi T$. We calculate
all one-loop contributions  to mass terms and coupling constants of the
static fields to order $g^{4}$, where $g$ denotes a gauge or top Yukawa coupling.
The second stage of reduction corresponds to the integration of heavy particles with
masses of the order of $g_{w} T$, $g_{s} T$.

\subsection{First Stage}

The potential in the 3D effective theory after integration over non-zero
Matsubara modes is of the form

\begin{eqnarray}
V &=& m_{H_{3}}^{2} H^{\dagger}H + \lambda_{H_{3}} (H^{\dagger}H)^2
+ m_{U_{3}}^{2} U^{\dagger}U + \lambda_{U_{3}} (U^{\dagger}U)^2 
+ \gamma_{3} (H^{\dagger}H)(U^{\dagger}U)\nonumber \\
&+& m_{Q_{3}}^{2} Q^{\dagger}Q  + m_{D_{3}}^{2} D^{\dagger}D 
+ \Lambda_{3}^{Q} (H^{\dagger}H)(Q^{\dagger}Q) + 
\Lambda_{4}^{c}  (H^{\dagger}Q)(Q^{\dagger}H)  \nonumber \\
&+& (\Lambda_{4}^{s} +
h_{t}^{L})|\epsilon_{ij} H^{i}Q^{j}|^{2}
+ (h_{t}^{QU} +
g_{s_{1}}^{QU})Q_{i\alpha}^{*}U_{\alpha}^{*}Q_{i\beta}U_{\beta}\nonumber \\
&+& g_{s_{2}}^{QU} U_{\alpha}U_{\alpha}^{*}Q_{j\gamma}^{*}Q_{j\gamma}
+ g_{s_{1}}^{QD} D_{\alpha}D_{\beta}^{*}Q_{j\beta}^{*}Q_{j\alpha}\nonumber \\
&+&  g_{s_{2}}^{QD} D_{\alpha}D_{\alpha}^{*}Q_{j\gamma}^{*}Q_{j\gamma}
+  g_{s_{1}}^{UD} U_{\alpha}U_{\gamma}^{*}D_{\gamma}^{*}D_{\alpha}\nonumber \\
& +& g_{s_{2}}^{UD} U_{\alpha}U_{\alpha}^{*}D_{\gamma}^{*}D_{\gamma}
+ \Lambda_{1} (Q^{\dagger}Q)^{2} +  \lambda_{D_{3}} (D^{\dagger}D)^{2}
+ \lambda_{Q_{3}}(Q_{i}^{\dagger}Q_{i})^{2}  \nonumber \\
& +& g_{s_{1}}^{QQ}
Q_{i\alpha}^{*}Q_{j\alpha}^{*}Q_{i\gamma}Q_{j\gamma}
+ g_{s_{2}}^{QQ} Q_{i\alpha}Q_{i\alpha}^{*}Q_{j\gamma}^{*}Q_{j\gamma} \nonumber \\
&+& {1\over 2} m_{A_{0}}^{2} A_{0}^{a}A_{0}^{a}  + {1\over 2} m_{C_{0}}^{2} C_{0}^{A}C_{0}^{A}
+ {1\over 4} g_{w_{3}}^{2} (H^{\dagger}H)(A_{0}^{a}A_{0}^{a}) \nonumber \\
&+& {1\over 4} g_{s_{3}}^{2} C_{0}^{A} C_{0}^{B} (U^{*})^{\dagger}\lambda^{A}\lambda^{B}U^{*}.
\label{3dtreepot}
\end{eqnarray}
Here $H$ is the Higgs doublet field, $U(D)$ is the right handed stop(sbottom) field, and $Q$ is the third generation left squark doublet field. The longitudinal components of the SU(2) and SU(3) gauge fields are denoted by $A_{o}$ and $C_{o}$, respectively. The latin (greek) indices indicate
SU(2) (SU(3)) components. As usual, the fields in eq. (\ref{3dtreepot}) are the static
components of the scalar fields properly renormalized, the dimension of the
fields in 3D is [GeV]$^{1/2}$. Quartic couplings are of order $g_{i}^{2}(h_{t}^{2}) T$, having dimensions of [GeV]; here $g_{i}(h_{t})$ denotes a gauge  ( top Yukawa) coupling.
In the following we have not included the correction to the quartic coupling 
between the doublet Higgs field and the triplet scalar field $A_{0}$, or the
corresponding correction for the SU(3) counterparts.
We work throughout in the Landau gauge. For an analysis of the gauge dependence,
we refer the reader to \cite{Laine}.

We present the full relations between 3D coupling constants and masses in
terms of the underlying 4D parameters and the temperature. Partial results for the MSSM in
the large-$m_{A}$ limit can be found in ref. \cite{Laine2} \footnote{If a
light higgsino is included, there will be important effects that
are proportional to the top Yukawa coupling in the dimensionally reduced
theory \cite{Losada1}.}.

\subsubsection{Mass terms}

For the  Higgs doublet we have\footnote{We mostly neglect the hypercharge coupling $g'$,
throughout the paper. The only exception is in the contribution to the tree-level
expression of the Higgs self-coupling $\lambda$, as this latter quantity is fundamental
in determining the strength of the phase transition.} 

\begin{eqnarray}
m_{H_{3}}^{2} &=& m_{H}^{2}\biggl(1 + {9\over 4} g_{w}^{2} {L_{b}\over 16\pi^{2}}
 - 3 h_{t}^{2} {L_{f}
\over 16\pi^{2}}\biggr)\nonumber \\
 &+&T^{2}\biggl({\lambda\over 2} + {3\over 16} g_{w}^{2}
 + {1\over 16} g'^{2} 
+ {1\over 4} h_{t}^{2} + {1\over 4}
(2h_{t}^{2}\sin^{2}\beta + 2\lambda_{3} + \lambda_{4})\biggr)\nonumber\\
&-&{L_{b}\over 16\pi^{2}}\biggl(6 \lambda m_{H}^{2}+ 3(m_{Q}^{2} + m_{U}^{2})h_{t}^{2}\sin^{2}\beta\biggr),
\label{mmH3}
\end{eqnarray}
where the Higgs mass parameter is denoted by $m_{H}$,
 and $\lambda = {(g_{w}^{2} + g'^{2})\over 8}\cos 2\beta, \lambda_{3} = {g_{w}^{2} \over 4}, \lambda_{4} = -{g_{w}^{2}\over 2}$, 
$L_{b}=2 \log{\overline{\mu} e^{\gamma}\over 4\pi T} \approx 2 \log{\mu\over 7.055 T}$, $L_{f} = L_{b} + 4\log2$.  Here $\overline{\mu}$ is the mass scale
defined by the modified minimal substraction scheme ($\overline{MS}$) scheme. Similarly, for the third generation squark mass terms we have

\begin{eqnarray}
m_{U_{3}}^{2} &=& m_{U}^{2}\biggl(1 + 4 g_{s}^{2} {L_{b}\over 16\pi^{2}}\biggr) +
 T^{2} \biggl({1\over 3} g_{s}^{2}
 + {2\over 3} \lambda_{U} + {1\over 6} h_{t}^{2}\sin^{2}\beta +
{1\over 6}h_{t}^{2} \biggr)
\nonumber \\
&-& {L_{b}\over 16 \pi^{2}}\biggl({4\over 3} g_{s}^{2} m_{U}^{2} + 2 h_{t}^{2} \sin^{2}\beta( m_{H}^{2}+
m_{Q}^{2}) \biggr),
\label{mmU3}
\end{eqnarray}

\begin{eqnarray}
m_{Q_{3}}^{2}& =& m_{Q}^{2}\biggl(1 + ({9\over 4} g_{w}^{2} 
 + 4 g_{s}^{2}){L_{b}\over 16\pi^{2}}\biggr) + T^{2}\biggl({3\over 16} g_{w}^{2}
+{ \lambda_{1}\over 2} + {4\over 9} g_{s}^{2}
 + {1\over 12} h_{t}^{2}(1 + \sin^{2}\beta)\biggr) \nonumber \\
&-& {L_{b}\over 16\pi^{2}}\biggl({4\over 3} g_{s}^{2} m_{Q}^{2} + 6 \lambda_{1} m_{Q}^{2}
+ h_{t}^{2} m_{U}^{2} + h_{t}^{2}\sin^{2}\beta m_{H}^{2}\biggr),
\label{mQ3}
\end{eqnarray}

\begin{eqnarray}
m_{D_{3}}^{2} = m_{D}^{2}\biggl(1 + 4 g_{s}^{2}{L_{b}\over 16\pi^{2}}\biggr) +  T^{2}\biggl({4\over
9} g_{s}^{2}\biggr)
-{L_{b}\over 16\pi^{2}}\biggl({4\over 3} g_{s}^{2} m_{D}^{2}\biggr).
\label{mD3}
\end{eqnarray}
where the soft SUSY-breaking masses for the third generation left squark
doublet and the right handed sbottom are denoted by $m_{Q}$ and $m_{D}$ respectively, and $\lambda_{U} = {g_{s}^{2}\over 6}$, $\lambda_{1} = {g_{w}^{2}\over 8}$.
The longitudinal components of the SU(2) and SU(3)  gauge fields acquire thermal masses given by

\begin{equation}
m_{A_{0}}^{2} = g_{w}^{2}T^{2}\biggl({2\over 3} + {N_{f}\over 12} + {N_{sw}\over
6}\biggr),
\label{mAo}
\end{equation}

\begin{equation}
m_{C_{0}}^{2} = g_{s}^{2}T^{2}\biggl(1 + {N_{f}\over 12} + {N_{ss} \over 6}\biggr),
\label{mCo}
\end{equation}
respectively, where $N_{sw}=4, N_{ss}=4, N_{f} = 6$ \cite{Laine2}.

\subsubsection{Couplings}

The 3D gauge coupling expressions for a dimensionally reduced $SU(N)$ gauge theory
can be found in ref. \cite{Laine2}. We include them for
completeness

\begin{equation}
g_{3}^{2} = T g^{2}(\overline{\mu})\biggl[1 + {g^{2}(\overline{\mu})\over 48 \pi^{2}}\biggl((22N -
 N_{s}){L_{b}\over 2}
- N_{f}L_{f}+ N\biggr)\biggr].
\label{gg3}
\end{equation}
Here $N_{s}$ is the number of scalar fields in the fundamental
representation and $N_{f}$ is the number of fermions.

For the scalar quartic self-couplings, we have the following relations
arising from the diagrams that have been shown in refs. \cite{Laine, Losada1}
\footnote{For contributions arising from the rest of the supersymmetric particles
 and the inclusion of Yukawa couplings for the other
(s)quarks see ref. \cite{Losada1}.},

\begin{eqnarray}
\lambda_{H_{3}} &=& \lambda T\biggl(1 + {9\over 2} g_{w}^{2}{ L_{b}\over 16\pi^{2}} -
6 h_{t}^{2} \sin^{2}\beta {L_{f}\over 16 \pi^2}\biggr) \nonumber \\
&-&T \biggl[ {L_{b}\over 16 \pi^{2}}\biggl( {9\over 16}g_{w}^4 + 
12 \lambda^{2} + 3\bigl( \lambda_{3}^{2} + \lambda_{3}\lambda_{4}
 + \lambda_{4}^{2}\cos^{4}\beta
+ \lambda_{4}^{2}\sin^{4}\beta  \nonumber \\
&+& h_{t}^{2}\sin^{2}\beta\bigl( \lambda_{3} + \lambda_{4} \sin^{2}\beta \bigr)
 + h_{t}^{4}\sin^{4}\beta \bigr) \biggr)  
+ {3\over 8}{g_{w}^{4}\over 16 \pi^{2}} 
+ 3 h_{t}^{4}\sin^{4}\beta {L_{f}\over 16\pi^{2}} \biggr],
\label{lalamH3}
\end{eqnarray}

\begin{eqnarray}
\lambda_{U_{3}} &=& {g_{s}^{2}\over 6} T\biggl(1 + 8 g_{s}^{2} {L_{b}\over
16\pi^{2}}\biggr)
-T\biggl[{L_{b}\over
16\pi^{2}}\biggl({23\over 36} g_{s}^{4} + {13\over 12} g_{s}^{4} - {2\over 3}
 h_{t}^{2}g_{s}^{4}\nonumber \\
& +& h_{t}^{4} +
h_{t}^{4}\sin^{4}\beta\biggl) + {13\over 18} g_{s}^{4}\biggr],
\label{lalamU3}
\end{eqnarray}

\begin{eqnarray}
\gamma_{3} &= &h_{t}^{2}\sin^{2}\beta T\biggl(1+ {9\over 4} g_{w}^{2} {L_{b}\over 16\pi^{2}}
 -3 h_{t}^{2}\sin^{2} 
{L_{f}\over 16\pi^{2}} + 4 g_{s}^{2} {L_{b}\over 16\pi^{2}}\biggr) \nonumber \\
&-&T\biggl[{L_{b}\over 16\pi^{2}}\biggl({4\over 3} h_{t}^{2} \sin^{2}\beta g_{s}^{2}
+ 2 h_{t}^{4}\sin^{4}\beta + 6 \lambda h_{t}^{2}\sin^{2}\beta\nonumber \\
&+& h_{t}^{2}(2\lambda_{3} +
\lambda_{4} + h_{t}^{2}\sin^{2}\beta\biggr)\biggr],
\label{gagamma3}
\end{eqnarray}

\begin{eqnarray}
\Lambda_{3}^{Q}& =& \lambda_{3}T\biggl(1 + {9\over 2} g_{w}^{2} {L_{b}\over 16\pi^{2}}
 -3 h_{t}^{2}\sin^{2} 
{L_{f}\over 16\pi^{2}} + 4 g_{s}^{2} {L_{b}\over 16\pi^{2}}\biggr) \nonumber \\
&-& T\biggl[{L_{b}\over 16\pi^{2}}\biggl({9\over 8} g_{w}^{4} + {4\over 3} g_{s}^{2} \lambda_{3} + 6 \lambda \lambda_{3}
+ 6 \lambda_{1}\lambda_{3} + 2 \lambda_{3}^{2} + 2 \lambda\lambda_{4}
\nonumber \\
&+& 2\lambda_{1}\lambda_{4}
+ \lambda_{4}^{2} \cos^{4}\beta + h_{t}^{4}\sin^{2}\beta + 2\lambda h_{t}^{2} \sin^{2}\beta
+ 2 \lambda_{1} h_{t}^{2}\sin^{2}\beta \nonumber \\
&+& h_{t}^{4}\sin^{4}\beta + 2 h_{t}^{2} \lambda_{4}
\sin^{4}\beta + \lambda_{4}^{2}\sin^{4}\beta\biggr)
 - {L_{f}\over 16\pi^{2}}2 h_{t}^{4}\sin^{4}\beta\biggr],
\label{l3Ql}
\end{eqnarray}

\begin{eqnarray}
\Lambda_{4}^{c}& =& \lambda_{4}\cos^{2}\beta T\biggl(1 + {9\over 2} g_{w}^{2} {L_{b}\over 16\pi^{2}} -3 h_{t}^{2}\sin^{2} 
{L_{f}\over 16\pi^{2}} + 4 g_{s}^{2} {L_{b}\over 16\pi^{2}}\biggr) \nonumber \\
&-& T{L_{b}\over 16\pi^{2}}\biggl( 2 \lambda\lambda_{4} \cos^{2}\beta + 2\lambda_{1}\lambda_{4}
\cos^{2}\beta + g_{w}^{2} \lambda_{4}\cos^{2}\beta
+ 2 \lambda_{4}^{2} \cos^{4}\beta\nonumber \\
&+& 4\lambda_{3}\lambda_{4}\cos^{2}\beta
+ {4\over 3} g_{s}^{2} \lambda_{4}\cos^{2}\beta \biggr),
\label{l4cb}
\end{eqnarray}

\begin{eqnarray}
\Lambda_{4}^{s}& =& \lambda_{4}\sin^{2}\beta T\biggl(1 + {9\over 2} g_{w}^{2} {L_{b}\over 16\pi^{2}} -3 h_{t}^{2}\sin^{2} 
{L_{f}\over 16\pi^{2}} + 4 g_{s}^{2} {L_{b}\over 16\pi^{2}}\biggr) \nonumber \\
&-& {T\over 2}{L_{b}\over 16\pi^{2}}\biggl({4\over 3} g_{s}^{2}(h_{t}^{2} +
 \lambda_{4}) \sin^{2}\beta
 + 2 \lambda\lambda_{4}\sin^{2}\beta + 2\lambda_{1}\lambda_{4}\sin^{2}\beta
+2 \lambda_{4}^{2} \sin^{4}\beta\nonumber \\
&+& 2h_{t}^{4}\sin^{4}\beta + 2\lambda h_{t}^{2} \sin^{2}\beta
+ 2 \lambda_{1} h_{t}^{2}\sin^{2}\beta \nonumber \\
&+& h_{t}^{4}\sin^{4}\beta + 4 h_{t}^{2} \lambda_{4}\sin^{4}\beta
 + g_{w}^{2} (h_{t}^{2} + \lambda_{4})\sin^{2}\beta
 +2 \lambda_{4}^{2}\sin^{4}\beta \nonumber \\
&+& 4\lambda_{3}\lambda_{4}\sin^{2}\beta
+ 4h_{t}^{2} \lambda_{3}\sin^{2}\beta \biggr),
\label{l4sb}
\end{eqnarray}

\begin{eqnarray}
h_{t}^{L} &=& h_{t}^{2} \sin^{2}\beta T\biggl(1 + {9\over 2} g_{w}^{2} {L_{b}\over 16\pi^{2}} -3 h_{t}^{2}\sin^{2} 
{L_{f}\over 16\pi^{2}} + 4 g_{s}^{2} {L_{b}\over 16\pi^{2}}\biggr) \nonumber \\
&-& {T\over 2}{L_{b}\over 16\pi^{2}}\biggl({4\over 3} g_{s}^{2}(h_{t}^{2} +
 \lambda_{4}) \sin^{2}\beta
 + 2 \lambda\lambda_{4}\sin^{2}\beta + 2\lambda_{1}\lambda_{4}\sin^{2}\beta
+2 \lambda_{4}^{2} \sin^{4}\beta\nonumber \\
&+& 2h_{t}^{4}\sin^{4}\beta + 2\lambda h_{t}^{2} \sin^{2}\beta
+ 2 \lambda_{1} h_{t}^{2}\sin^{2}\beta \nonumber \\
&+& h_{t}^{4}\sin^{4}\beta + 4 h_{t}^{2} \lambda_{4}\sin^{4}\beta
 + g_{w}^{2} (h_{t}^{2} + \lambda_{4})\sin^{2}\beta
 +2 \lambda_{4}^{2}\sin^{4}\beta \nonumber \\
&+& 4\lambda_{3}\lambda_{4}\sin^{2}\beta
+ 4h_{t}^{2} \lambda_{3}\sin^{2}\beta \biggr).
\label{htL}
\end{eqnarray}

 We include the relations for the couplings among the heavy fields which will be integrated out at
the second stage. These relations are needed only when the 2-loop contribution from these fields to the effective potential are included, see section 2.4.
We obtain
\begin{eqnarray}
g_{s_{1}}^{QU} &=& -{1\over 2} g_{s}^{2}T\biggl(1 + \biggl({9\over 4} g_{w}^{2}
 + 8 g_{s}^{2}\biggr){L_{b}\over 16\pi^{2}}\biggr)
-{T\over 2}\biggl({L_{b}\over 16\pi^{2}}\biggl(3 h_{t}^{4} + {5\over 4} g_{s}^{4}\nonumber \\
&-&  {7\over 6} h_{t}^{2} g_{s}^{2} - {5\over 12} g_{s}^{4}
+ {3\over 8} g_{w}^{2}(2 h_{t}^{2} -g_{s}^{2})\biggr) 
+ {5\over 6}{g_{s}^{4}\over 16\pi^{2}}\biggr),
\label{gsQU1}
\end{eqnarray}

\begin{eqnarray}
h_{t}^{QU}& =& h_{t}^{2}T\biggl(1 + \biggl({9\over 4} g_{w}^{2} +
 8 g_{s}^{2}\biggr){L_{b}\over 16\pi^{2}}\biggr)
-{T\over 2}\biggl({L_{b}\over 16\pi^{2}}\biggl(3 h_{t}^{4} + {5\over 4} g_{s}^{4}\nonumber \\
 &-& {7\over 6} h_{t}^{2} g_{s}^{2} - {5\over 12} g_{s}^{4}
+ {3\over 8} g_{w}^{2}(2 h_{t}^{2} -g_{s}^{2})\biggr)
 + {5\over 6}{g_{s}^{4}\over 16\pi^{2}}\biggr),
\label{htQU}
\end{eqnarray}

\begin{eqnarray}
g_{s_{2}}^{QU}& =& {1\over 6} g_{s}^{2}T\biggl(1 + \biggl({9\over 4} g_{w}^{2} +
 8 g_{s}^{2}\biggr){L_{b}\over 16\pi^{2}}\biggr)
-T\biggl({L_{b}\over 16\pi^{2}}\biggl( h_{t}^{4} +  g_{s}^{2}\lambda_{1}\nonumber \\
 &-&  {1\over 2} h_{t}^{2} g_{s}^{2} 
+{7\over 12} g_{s}^{4} + 2h_{t}^{2}\lambda_{3}\sin^{2}\beta + h_{t}^{2}\lambda_{4}\sin^{2}\beta
\cos^{2}2\beta \nonumber \\
& +& {11\over 12} g_{s}^{4} + h_{t}^{4}\sin^{4}\beta + h_{t}^{2}\lambda_{4}\sin^{4}\beta\biggr)
 + {11\over 18}{g_{s}^{4}\over 16\pi^{2}}\biggr),
\label{gsQU2}
\end{eqnarray}

\begin{equation}
g_{s_{1}}^{UD} = {1\over 2} g_{s}^{2}T\biggl(1 +
  8 g_{s}^{2}{L_{b}\over 16\pi^{2}}\biggr)\nonumber \\
- T\biggl({L_{b}\over 16\pi^{2}}\biggl(- h_{t}^{2}g_{s}^{2} + {5\over 2}
g_{s}^{4}\biggr)
 + {5\over
6} {g_{s}^{4}
\over 16\pi^{2}}\biggr),
\label{gsDU1}
\end{equation}

\begin{eqnarray}
g_{s_{2}}^{UD} &=& -  {1\over 6} g_{s}^{2}T\biggl(1 + 
 8 g_{s}^{2}{L_{b}\over 16\pi^{2}}\biggr)\nonumber \\
&-& T\biggl({L_{b}\over 16\pi^{2}}\biggl({1\over 3} h_{t}^{2}g_{s}^{2} + {1\over 36} g_{s}^{4} 
+  {11\over 12} g_{s}^{4}\biggr) + {11\over 18} {g_{s}^{4}
\over 16\pi^{2}}\biggr),
\label{gsDU2}
\end{eqnarray}

\begin{eqnarray}
g_{s_{1}}^{QD}& =& -{1\over 2} g_{s}^{2}T\biggl(1 + \biggl({9\over 4} g_{w}^{2}+
 8 g_{s}^{2}\biggr){L_{b}\over 16\pi^{2}}\biggr)
- T\biggl({L_{b}\over 16\pi^{2}}\biggl( h_{t}^{2}g_{s}^{2} + {5\over 4} g_{s}^{4}\nonumber \\
 &-& {5\over 12} g_{s}^{4} 
+ {3\over 8} g_{w}^{2}(2 h_{t}^{2} -g_{s}^{2})\biggr) + {5\over 6} {g_{s}^{4}
\over 16\pi^{2}}\biggr),
\label{gsQD1}
\end{eqnarray}

\begin{eqnarray}
g_{s_{2}}^{QD}& =& {1\over 6} g_{s}^{2}T\biggl(1 + \biggl({9\over 4} g_{w}^{2}+
 8 g_{s}^{2}\biggr){L_{b}\over 16\pi^{2}}\biggr)
- T\biggl({L_{b}\over 16\pi^{2}}\biggl( -{1\over 6} h_{t}^{2}g_{s}^{2} + {11\over 12}
 g_{s}^{4}\nonumber \\
& +&{7\over 12} g_{s}^{4}  +
g_{s}^{2}\lambda_{1}
+ {3\over 8} g_{w}^{2}(2 h_{t}^{2} -g_{s}^{2})\biggr) + {11\over 18} {g_{s}^{4}
\over 16\pi^{2}}\biggr),
\label{gsQD2}
\end{eqnarray}

\begin{eqnarray}
\Lambda_{1} & = & \lambda_{1}T\biggl(1+ \biggl({9\over 2}g_{w}^{2} + 8g_{s}^{2}\biggr){L_{b}\over
(16 \pi^{2})}\biggr) - {T\over 2}\biggl({L_{b}\over
(16 \pi^{2})}\biggl(h_{t}^{4} -{2\over 3} h_{t}^{2} g_{s}^{2} \nonumber \\
&+& {23\over 18} g_{s}^{4} + {5\over 16} g_{w}^{4} + {3\over 4} g_{s}^{2} g_{w}^{2}
+ 2\lambda_{3}^{2} + 2 \lambda_{3}\lambda_{4} + \lambda_{4}^{2}\cos^{4}\beta + 2 h_{t}^{2}\lambda_{3} \sin^{2}\beta  \nonumber \\
& +& h_{t}^{4}\sin^{4}\beta + 2 h_{t}^{2}\lambda_{4}\sin^{4}\beta
+ \lambda_{4}^{2}\sin^{4}\beta +{13\over 12}g_{s}^{4}\biggr) + {13\over 18}
g_{s}^{4}\biggr),
\label{Lam1}
\end{eqnarray}

\begin{eqnarray}
\lambda_{Q_{3}} &=& {g_{s}^{2}\over 6}T\biggl(1+ \biggl({9\over 2}g_{w}^{2} + 8g_{s}^{2}\biggr){L_{b}\over 
(16 \pi^{2})}\biggr) - {T\over 2}\biggl({L_{b}\over
(16 \pi^{2})}\biggl(h_{t}^{4} -{2\over 3} h_{t}^{2} g_{s}^{2} \nonumber \\
&+& {23\over 18} g_{s}^{4} + {5\over 16} g_{w}^{4} + {3\over 4} g_{s}^{2} g_{w}^{2}
+ 2\lambda_{3}^{2} + 2 \lambda_{3}\lambda_{4} + \lambda_{4}^{2}\cos^{4}\beta + 2 h_{t}^{2}\lambda_{3} \sin^{2}\beta + h_{t}^{4}\sin^{4}\beta \nonumber \\
& +& 2 h_{t}^{2}\lambda_{4}\sin^{4}\beta
+ \lambda_{4}^{2}\sin^{4}\beta +{13\over 12}g_{s}^{4}\biggr) + {13\over 18}
g_{s}^{4}\biggr),
\label{lamQ3}
\end{eqnarray}

\begin{eqnarray}
g_{s_{1}}^{QQ}& =& {g_{s}^{2}\over 6}T\biggl(1+ \biggl({9\over 2}g_{w}^{2} + 8g_{s}^{2}\biggr){L_{b}\over
(16 \pi^{2})}\biggr) - T\biggl({L_{b}\over
(16 \pi^{2})}\biggl(h_{t}^{4} - h_{t}^{2} g_{s}^{2}\nonumber \\
& +& {11\over 12}g_{s}^{4}
+ {5\over 4 }g_{s}^{4} + \lambda_{4}^{2}\cos^{2}\beta - 2 h_{t}^{2}\lambda_{4}
\sin^{2}\beta \cos^{2}\beta - 2\lambda_{4}^{2}\sin^{2}\beta \cos^{2}\beta
+ h_{t}^{4}\nonumber \\
& +& 2 h_{t}^{2}\lambda_{4}\sin^{4}\beta + \lambda_{4}^{2}\sin^{2}\beta + g_{s}^{2}g_{w}^{2} - {3\over 16} g_{w}^{4}\biggr) + {5\over 6} {g_{s}^{4}\over (16\pi^{2})}\biggr),
\label{gs1QQ}
\end{eqnarray}

\begin{eqnarray}
g_{s_{2}}^{QQ}& =& {g_{s}^{2}\over 6}T(1+ \biggl({9\over 2}g_{w}^{2} + 8g_{s}^{2}\biggr){L_{b}\over
(16 \pi^{2})}\biggr) - {T\over 2}\biggl({L_{b}\over
(16 \pi^{2})}\biggl({1\over 3}h_{t}^{2} g_{s}^{2} + {1\over 36} g_{s}^{4} \nonumber \\
&+& {11\over 12} g_{s}^{4} + 2\lambda_{3}^{2} + 2\lambda_{3}\lambda_{4} + 2 h_{t}^{2}\lambda_{3} \sin^{2}\beta + 2 h_{t}^{2}\lambda_{4}\sin^{2}\beta \cos^{2}\beta\nonumber \\
&+& 2 \lambda_{4}^{2}\sin^{2}\beta\cos^{2}\beta + {1\over 8}g_{w}^{4} -{1\over 4} g_{s}^{2}g_{w}^{2}\biggr) + {11\over 18}{g_{s}^{4}\over (16\pi^{2})}\biggr).
\label{gs2QQ}
\end{eqnarray}

A few  technical comments are in order. We  point out  that the full one-loop contribution
to the quartic coupling  $|\epsilon_{ij}H^{i}Q^{j}|^{2}$ is given by the sum of eqs. (\ref{l4sb}) and (\ref{htL}). Similarly, for the quartic coupling
$Q_{i\alpha}^{*}U_{\alpha}^{*}Q_{i\beta}U_{\beta}$ the full contributions
arises from eqs. (\ref{gsQU1}) and (\ref{htQU}). 
 It is also important to note that there are off-diagonal (in colour space) gluonic contributions  to the quartic couplings involving the strong gauge coupling.

Additionally, when the full supersymmetric spectrum is not included then the running of each of the strong quartic couplings given in eqs.
(\ref{gsQU1})-(\ref{gs2QQ}) is different. Although 
 the
gluino contribution is decoupled under our assumptions, we now write as a check  the
gluino contributions to the logarithmic part of the quartic couplings.
This shows how, if one includes the full spectrum, then the relation
between the beta-function coefficients is the same
as the relation between the couplings.

The gluino contribution, in units of 
 $T {L_{f}\over 16\pi^{2}} {1\over 3} g_{s}^{4}$, to the self-couplings is
${22\over 3}$.
Using  the same units, for $g_{s_{1}}^{QU}, g_{s_{1}}^{QD}$ the contribution is $-4$,
for $g_{s_{2}}^{QU}, g_{s_{2}}^{QD}$ the contribution is ${20\over 3}$,
for $g_{s_{1}}^{UD}, g_{s_{1}}^{QQ}$ the additional term is  $14$, and for $g_{s_{2}}^{UD} , g_{s_{2}}^{QQ}$ the contribution is ${2\over 3}$.

\subsection{Second Stage}

 Another
simplification of the effective theory can be obtained by integrating out the
scalar fields which are massive at the transition point.
As we have seen the static modes corresponding to the scalar fields $Q, D, A_{o}, C_{o}$, acquired thermal masses proportional to $\sim g_{w(s)}T$, as a consequence of the integration out of the non-zero Matsubara modes.
The second stage proceeds in exactly the same way as in reference \cite{Laine2}.
We include the additional corrections arising from the couplings we have considered.

\subsubsection{Couplings}
The final expression for the tree level 3D potential is given by

\begin{equation}
V_{3D} = \overline{m}_{H_{3}}^{2} H^{\dagger}H + \overline{\lambda}_{H_{3}}(H^{\dagger}H)^{2} + \overline{m}_{U_{3}}^{2} U^{\dagger}U + \overline{\lambda}_{U_{3}}(U^{\dagger}U)^{2}
+ \overline{\gamma}_{3} H^{\dagger}H U^{\dagger}U,
\label{v3D}
\end{equation}
where the scalar couplings are now

\begin{eqnarray}
\overline{\lambda}_{H_{3}} &=& \lambda_{H_{3}} - {3\over 16} {g_{w_{3}}^{4}\over 8\pi m_{A_{0}}}
-{3\over 8\pi m_{Q_{3}}}\biggl(\Lambda_{3}^{2} + \Lambda_{3}(\Lambda_{4}^{c} + 
\Lambda_{4}^{s}) \nonumber \\
&+& {1\over 2} \biggl((\Lambda_{4}^{c})^{2} + (\Lambda_{4}^{s})^{2}\biggr) + h_{t}^{L} \Lambda_{3}^{Q}
+ h_{t}^{L}\Lambda_{4}^{s} + {1\over 2}( h_{t}^{L})^{2}\biggr),
\label{lamH3}
\end{eqnarray}

\begin{eqnarray}
\overline{\lambda}_{U_{3}} &=& \lambda_{U_{3}} -
 {13\over 36} {g_{s_{3}}^{4}\over 8\pi m_{C_{0}}}
-{1\over 8\pi m_{D_{3}}}\biggl({1\over 2}(g_{s_{1}}^{UD} + g_{s_{2}}^{UD})^{2}
 + (g_{s_{2}}^{UD})^{2}\biggr) \nonumber \\
&-& {1\over 8\pi m_{Q_{3}}}\biggl((h_{t}^{QU})^{2} - 2 h_{t}^{QU} g_{s_{1}}^{QU} + 
2h_{t}^{QU} g_{s_{2}}^{QU} +(g_{s_{1}}^{UD} + g_{s_{2}}^{UD})^{2} +
2(g_{s_{2}}^{UD})^{2}\biggr),
\label{lamU3}
\end{eqnarray}

\begin{eqnarray}
\overline{\gamma_{3}} &=& \gamma_{3} -{1\over 8\pi m_{Q_{3}}}(h_{t}^{QU} + g_{s_{1}}^{QU}+ 3 g_{s_{2}}^{QU})(2\Lambda_{3}^{Q}
+ \Lambda_{4}^{c} + \Lambda_{4}^{s} + h_{t}^{L}).
\label{gamma3}
\end{eqnarray}

The 3D gauge couplings which appear in the SU(2) and SU(3) covariant
derivatives of the effective theory are

\begin{equation}
\overline{g}_{w_{3}}^{2} = g_{w_{3}}^{2}\biggl(1 - {g_{w_{3}}^{2}\over 24 \pi m_{A_{0}}} 
- {g_{w_{3}}^{2}\over 16\pi m_{Q_{3}}}\biggr),
\label{gw3}
\end{equation}

\begin{equation}
\overline{g}_{s_{3}}^{2} = g_{s_{3}}^{2}\biggl(1 - {g_{s_{3}}^{2}\over 16 \pi m_{C_{0}}} 
- {g_{s_{3}}^{2}\over 24\pi m_{Q_{3}}} - {g_{s_{3}}^{2} \over 48\pi m_{D_{3}}}\biggr).
\label{gs3}
\end{equation}

\subsubsection{Mass terms}

The one-loop contribution to the mass terms can be obtained directly as shown in
ref. \cite{KLRS}:

\begin{eqnarray}
\overline{m}_{H_{3}}^{2} &=& m_{H_{3}}^{2} - {3\over 16\pi} g_{w_{3}} m_{A_{0}} -{3\over 4\pi}
(2\Lambda_{3}^{Q} +  \Lambda_{4}^{c} + \Lambda_{4}^{s} + h_{t}^{L})m_{Q_{3}},
\label{mH3}
\end{eqnarray}

\begin{eqnarray}
\overline{m}_{U_{3}}^{2} &=& m_{U_{3}}^{2} - {1\over 3\pi} g_{s_{3}} m_{C_{0}} -{1\over 4\pi}
(2 h_{t}^{QU} + 2 g_{s_{1}}^{QU} + 6 g_{s_{2}}^{QU})m_{Q_{3}}\nonumber \\
& -&{1\over 4\pi}(2g_{s_{1}}^{UD}
+ 6g_{s_{2}}^{UD})m_{D_{3}}.
\label{mU3}
\end{eqnarray}

Until now, our procedure has been exactly the same as  in previous 3D
reductions of the MSSM. However, in order to precisely fix the scales of the
couplings that appear in the thermal polarizations of eqs. (2)-(3), one needs to perform 
a 2-loop evaluation of the effective potential. In  their calculation,
 Bodeker et al. \cite {Laine2} took the values
of the couplings in the screening parts of
the 3D masses to be equal to the 3D values of
these couplings. This   is the correct result
when  two-loop corrections are included \cite{KLRS, oespinosa}, see eqs.  (\ref{mU3mu}) and (\ref{mH3mu})  below.
 In addition, the mass parameters are
renormalized in the 3D theory,

\begin{equation}
\overline{m}_{H_{3}}^{2}(\mu) = \overline{m}_{H_{3}}^{2} + {1\over (16\pi^{2})} f_{2m_{H}}\log{\Lambda_{H_{3}}\over \mu},
\label{mH3mu2}
\end{equation}

\begin{equation}
\overline{m}_{U_{3}}^{2}(\mu) = \overline{m}_{U_{3}}^{2} + {1\over (16\pi^{2})} f_{2m_{U}}\log{\Lambda_{U_{3}}\over \mu}.
\label{mU3mu2}
\end{equation}
The expressions for the 2-loop beta functions $f_{2m_{H}},f_{2m_{U}}$ for the mass parameters
have been given in ref. \cite{Laine2}. As mentioned there
in order to fix the values of the parameters $\Lambda_{H_{3}}$ and $\Lambda_{U_{3}}$ we must employ  the 2-loop  effective potential of the 4D theory.
In refs. \cite{Laine2, Laine3} an estimate of $\Lambda_{H_{3}} \sim
\Lambda_{U_{3}} \sim 7T$ was used.
The expressions for the 2-loop effective potential
in a $H$($\phi$-direction) and $U$($\chi$-direction) background have been given
for 4D in the paper by Carena et al. \cite{Carena2}, and for 3D by Bodeker et al.
\cite{Laine}. In sections 2.3 and 2.4 we perform a two-loop calculation
of the effective potential, incorporating all of the
corrections to the 3D couplings obtained in the previous sections, to determine the exact values of
$\Lambda_{H_{3}}$ and $\Lambda_{U_{3}}$.

We will analyse the effect on the critical temperatures when these
corrections are included. Qualitatively we can
say that if the net effect  increases the value of $\Lambda_{H_{3}}$,  then the critical temperature in the $\phi$-direction
decreases, and vice versa. A similar effect occurs in the $\chi$-direction. Thus,
 as the range of values of $m_{\tilde{t}_{R}}$, which has been determined
to give rise to a two-stage phase transition, is
small, a more precise
evaluation of  the critical temperatures
is of interest. This could have the effect of either reducing or enhancing
the allowed range of values of the right stop mass, $m_{\tilde{t}_{R}}$, for which
a two-stage phase transition can occur.

\subsection{Two-loop contributions}

The strategy we employ follows that of ref. \cite{KLRS}.
The idea is to use the 4D 2-loop effective potential in order to fix
the scales in the 3D theory, and to use the 3D effective potential expressions
for the Higgs and stop fields
given in ref. \cite{Laine2} to analyse the phase transition.
We  calculate the unresummed 2-loop effective potential
in order to include all 4D corrections to the mass parameters, resummation
is automatically  included in the calculation of the 2-loop effective
potential in the 3D theory. We must also include the
contributions  to the 2-loop effective potential of the static modes, which have been
integrated out  at the second stage  (includes the effects of resummation of the heavy fields).

There are several effects that must be considered in order to obtain
all of the contributions (constant and logarithmic) to the mass 
parameters. From the 4D effective potential one finds the two-loop contributions
from the gauge bosons, Higgs, right handed stop, left handed squark doublet, right handed sbottom,
and top quark. The expression for the 2-loop effective  potential can be found in refs. \cite{Espinosa, Laine2,
Carena2}. In particular, within our approximations for the $\phi$-direction,
the appropriate expression is that given in ref. \cite{Espinosa}, because
we are including  the 3rd generation squark doublet and the right handed sbottom
in the thermal bath\footnote{For most of our analysis we will include the third-generation left squark
doublet in the thermal bath.}. The main difference is that the $D$ functions appearing below
correspond to the unresummed
expressions. Additionally we
must include the effects arising  at the second
stage of reduction from  the left handed squark doublet, the right handed sbottom, the scalar  triplet and the scalar 
octet.
We now derive  the effective potential at finite temperature using  the
background fields  $\phi$ and $\chi= \tilde{t}_{R\alpha}u^{\alpha}$, where
we have chosen the unit vector in colour space $u^{\alpha}=(1,0,0)$. 
We now write the expressions in the shifted theory of
 the mass spectrum after the first stage of integration. The gauge bosons masses are

\begin{equation}
m_{W,Z}^{2} = {1\over 4} g_{w}^{2} \phi^{2},\hspace{.2in} m_{G}^{2}= {1\over 4} g_{s}^{2} \chi^{2},
\hspace{.2in} \overline{m}_{G}^{2} = {4\over 3} m_{G}^{2}.
\label{gaugemasses}
\end{equation}
With no mixing in the Higgs sector, the Goldstone bosons and Higgs masses are 
\begin{eqnarray}
m_{\pi}^{2} &=& m_{H}^{2} + \lambda \phi^{2} + h_{t}^{2}\sin^{2}\beta {\chi^{2}\over 2}, \nonumber \\
m_{h}^{2} &=& m_{H}^{2} + 3\lambda \phi^{2} + h_{t}^{2}\sin^{2}\beta {\chi^{2}\over 2}, \nonumber \\
m_{\omega}^{2} &=& m_{\overline{\omega}}^{2} = m_{U}^{2} + \lambda_{U} \chi^{2} + h_{t}^{2}\sin^{2}\beta {\phi^{2}\over 2}, \nonumber \\
m_{u}^{2} &=& m_{U}^{2} + 3\lambda \chi^{2} + h_{t}^{2}\sin^{2}\beta {\phi^{2}\over 2}. 
\label{mpi}
\end{eqnarray}
The masses of the rest of the scalars contributing to the
effective potential are given by
\begin{equation}
m_{\tilde{t}_{L_{1}}}^{2} = m_{Q}^{2} + (h_{t}^{2}\sin^{2}\beta + \lambda_{3} + 
\lambda_{4}\sin^{2}\beta){\phi^{2}\over 2} + \biggl(h_{t}^{2} - {g_{s}^{2}\over 3}\biggr)
{\chi^{2}\over 2},
\label{mtL1}
\end{equation}

\begin{equation}
m_{\tilde{t}_{L_{2,3}}}^{2} = m_{Q}^{2} + (h_{t}^{2}\sin^{2}\beta + \lambda_{3} + 
\lambda_{4}\sin^{2}\beta){\phi^{2}\over 2} + \biggl({g_{s}^{2}\over 6}\biggr)
{\chi^{2}\over 2},
\label{mtL2}
\end{equation}

\begin{equation}
m_{\tilde{b}_{L_{1}}}^{2} = m_{Q}^{2} + ( \lambda_{3} + 
\lambda_{4}\cos^{2}\beta){\phi^{2}\over 2} + \biggl(h_{t}^{2} - {g_{s}^{2}\over 3}\biggr)
{\chi^{2}\over 2},
\label{mbL1}
\end{equation}

\begin{equation}
m_{\tilde{b}_{L_{2,3}}}^{2} = m_{Q}^{2} + (\lambda_{3} + 
\lambda_{4}\cos^{2}\beta){\phi^{2}\over 2} + \biggl({g_{s}^{2}\over 6}\biggr)
{\chi^{2}\over 2},
\label{mbL2}
\end{equation}

\begin{equation}
m_{\tilde{b}_{R_{1}}}^{2} = m_{D}^{2} + \biggl({g_{s}^{2}\over 6}\biggr)
{\chi^{2}\over 2},
\label{mbR1}
\end{equation}

\begin{equation}
m_{\tilde{b}_{R_{2,3}}}^{2} = m_{D}^{2}  - \biggl({g_{s}^{2}\over 3}\biggr)
{\chi^{2}\over 2}.
\label{mbR2}
\end{equation}

\subsubsection{$\phi$-direction}

The contributions to the terms in $m_{H_{3}}^{2}(\mu)$ proportional 
to $g_{w}^{4}$, $g_{w}^{2} \lambda_{H_{3}}$ and $\lambda_{H_{3}}^{2}$
from Standard Model particles
have been calculated in the paper by Kajantie et al. \cite{KLRS}. The correct
expression is obtained from eq. (151) of \cite{KLRS},
substituting $\lambda = {1\over 8} (g_{w}^{2} + g'^{2})\cos 2\beta$ and $g_{Y}^{2} =
h_{t}^{2}\sin^{2}\beta$. For the Standard Model,  this already includes the finite
contributions from counterterms.

The additional corrections that arise from supersymmetric particles can be calculated
using the 2-loop unresummed potential. The expressions in
the integral form have been given in ref. \cite{Espinosa}
for zero squark mixing, and
we include them for completeness. Our notation
for the $D$-functions  corresponds to that of ref. \cite{KLRS}.
The contributions from the 2-loop graphs containing supersymmetric particles are given below. For the $\phi$-direction,
we can drop the colour index of the squark masses:

\begin{eqnarray}
(SSV)& =& -{g_{w}^{2}\over 8} N_{c}[D_{SSV}(m_{\tilde{t}_{L}},m_{\tilde{t}_{L}},m_{W})
+ D_{SSV}(m_{\tilde{b}_{L}},m_{\tilde{b}_{L}},m_{W}) +
 4 D_{SSV}(m_{\tilde{t}_{L}},m_{\tilde{b}_{L}},m_{W})] \nonumber \\
&-&{g_{s}^{2}\over 4}(N_{c}^{2} -1)[D_{SSV}(m_{\tilde{t}_{L}},m_{\tilde{t}_{L}},0) +
D_{SSV}(m_{\tilde{b}_{L}},m_{\tilde{b}_{L}},0) \nonumber \\&+&
D_{SSV}(m_{\tilde{t}_{R}},m_{\tilde{t}_{R}},0)+
D_{SSV}(m_{\tilde{b}_{R}},m_{\tilde{b}_{R}},0)],
\label{SSVphi}
\end{eqnarray}

\begin{eqnarray}
(SSS)& =& -N_{c}\biggl[\biggl(h_{t}^{2} \sin^{2}\beta + {g_{w}^{2}\over 4}\cos 2\beta\biggr)^{2}
D_{SSS}(m_{\tilde{t}_{L}},m_{\tilde{t}_{L}},m_{h})
+ \biggl( {g_{w}^{2}\over 4}\cos 2\beta\biggr)^{2} D_{SSS}(m_{\tilde{b}_{L}},m_{\tilde{b}_{L}},m_{h})
\nonumber \\
&+& (h_{t}^{2} \sin^{2}\beta)^{2} D_{SSS}(m_{\tilde{t}_{R}},m_{\tilde{t}_{R}},m_{h})
+ \biggl(h_{t}^{2}\sin^{2}\beta+
 {g_{w}^{2}\over 2}\cos 2\beta\biggr)^{2}
D_{SSS}(m_{\tilde{t}_{L}},m_{\tilde{b}_{L}},m_{\pi})\biggr]{\phi^{2}\over 2},
\label{SSSphi}
\end{eqnarray}

\begin{eqnarray}
(SV)& =& -{1\over 4} g_{s}^{2}(N_{c}^{2} -1)[D_{SV}(m_{\tilde{t}_{L}},0) +
 D_{SV}(m_{\tilde{b}_{L}},0) + D_{SV}(m_{\tilde{t}_{R}},0) +
D_{SV}(m_{\tilde{b}_{R}},0)]\nonumber \\
&-& {3\over 8} g_{w}^{2}N_{c}[D_{SV}(m_{\tilde{t}_{L}},m_{W})+
D_{SV}(m_{\tilde{b}_{L}},m_{W})],
\label{SVphi}
\end{eqnarray}

\begin{eqnarray}
(SS) &=& {g_{w}^{2}\over 4} N_{c}(2- N_{c}) D_{SS}(m_{\tilde{t}_{L}},m_{\tilde{b}_{L}})+
h_{t}^{2}N_{c}[D_{SS}(m_{\tilde{t}_{L}},m_{\tilde{t}_{R}}) +
D_{SS}(m_{\tilde{b}_{L}},m_{\tilde{t}_{R}})] \nonumber \\
&+& \biggl({g_{w}^{2}\over 8} +{g_{s}^{2}\over 6}\biggr)N_{c}(N_{c}+
1)[D_{SS}(m_{\tilde{t}_{L}},m_{\tilde{t}_{L}}) +
D_{SS}(m_{\tilde{b}_{L}},m_{\tilde{b}_{L}})] \nonumber \\
&+& {g_{s}^{2}\over 6} N_{c}(N_{c}+1)[D_{SS}(m_{\tilde{t}_{R}},m_{\tilde{t}_{R}})
+ D_{SS}(m_{\tilde{b}_{R}},m_{\tilde{b}_{R}})] \nonumber \\
&+& N_{c}\biggl({1\over 2} h_{t}^{2}\sin^{2}\beta + {1\over 8}g_{w}^{2} \cos 2\beta\biggr)[
D_{SS}(m_{\tilde{t}_{L}},m_{h}) + 2D_{SS}(m_{\tilde{b}_{L}},m_{\pi})
+ D_{SS}(m_{\tilde{t}_{L}},m_{\pi})] \nonumber \\
&-& {1\over 8} N_{c}g_{w}^{2}\cos 2\beta[D_{SS}(m_{\tilde{b}_{L}},m_{h}) + 2
D_{SS}(m_{\tilde{t}_{L}},m_{\pi}) + D_{SS}(m_{\tilde{b}_{L}},m_{\pi})]\nonumber \\
&+& {1\over 2} N_{c} h_{t}^{2}\sin^{2}\beta[D_{SS}(m_{\tilde{t}_{R}},m_{h}) + 
3D_{SS}(m_{\tilde{t}_{R}},m_{\pi})].
\label{SSphi}
\end{eqnarray}

There are also counterterm contributions 
to the mass terms; for the $\phi$-direction
they correspond to eq. (B.3) of ref. \cite{Espinosa}.

\begin{equation}
\delta V = -{T^{2}\over 16\pi^{2}}{\phi^{2}\over 96}(3 g_{w}^{4} + {11\over 9}g'^{4}).
\label{countermphi}
\end{equation}

\subsubsection{ $\chi$-direction} 

The two-loop unresummed effective potential in the $\chi$-direction is given
by the following contri-\\butions\footnote{The expression given for the 4D effective potential
would correspond to the
usual resummed 2-loop 4D effective potential if we use 
the resummed expressions in the $D$ functions appearing below.}$^{,}$\footnote{As $m_{\tilde{t}_{L2}} = m_{\tilde{t}_{L3}}, m_{\tilde{b}_{L2}}=m_{\tilde{b}_{L3}}, m_{\tilde{b}_{R2}}= m_{\tilde{b}_{R3}}$ in the $\chi$-direction, we just multiply 
by a factor of 2 the contributions from these fields in some of the following expressions.}:

\begin{equation}
(VVV) = -g_{s}^{2}{ N_{c}\over 4}[(N_{c}-2) D_{VVV}(m_{G},m_{G},0)
+ D_{VVV}(m_{G},m_{G},\overline{m}_{G})],
\label{VVVchi}
\end{equation}

\begin{equation}
(\eta\eta V) = -g_{s}^{2}{N_{c}\over 2}[2(N_{c}-1) D_{\eta\eta V}(0,0,m_{G})+
D_{\eta\eta V}(0,0,\overline{m}_{G})],
\label{eeVchi}
\end{equation}

\begin{equation}
(VV) = -g_{s}^{2} {N_{c}\over 8}[2(N_{c} -2) D_{VV}(0,m_{G}) + 2 D_{VV}(m_{G},\overline{m}_{G})
+ (N_{c}-1) D_{VV}(m_{G},m_{G})],
\label{VVchi}
\end{equation}

\begin{eqnarray}
(SSV) &=& -{g_{w}^{2}\over 8}[D_{SSV}(m_{\tilde{t}_{L1}},m_{\tilde{t}_{L1}},0) +
D_{SSV}(m_{\tilde{t}_{L2}},m_{\tilde{t}_{L2}},0) +
D_{SSV}(m_{\tilde{t}_{L3}},m_{\tilde{t}_{L3}},0)\nonumber \\
& +&
D_{SSV}(m_{\tilde{b}_{L1}},m_{\tilde{b}_{L1}},0) +
D_{SSV}(m_{\tilde{b}_{L2}},m_{\tilde{b}_{L2}},0) +
D_{SSV}(m_{\tilde{b}_{L3}},m_{\tilde{b}_{L3}},0)\nonumber \\
& +&
4(D_{SSV}(m_{\tilde{t}_{L1}},m_{\tilde{b}_{L1}},0) + D_{SSV}(m_{\tilde{t}_{L2}},m_{\tilde{b}_{L2}},0) +
D_{SSV}(m_{\tilde{t}_{L3}},m_{\tilde{b}_{L3}},0) )]\nonumber \\
&-& g_{s}^{2} {1\over 4}[(N_{c} -1) D_{SSV}(m_{\omega},\overline{m}_{\omega},m_{G})
+ (N_{c}-1)  D_{SSV}(m_{\omega},m_{u},m_{G})\nonumber \\
& +& {N_{c}-1\over N_{c}}
 D_{SSV}(\overline{m}_{\omega},m_{u},m_{G}) + {1\over N_{c}}
D_{SSV}(m_{\omega},m_{\omega},\overline{m}_{G}) \nonumber \\
&+& N_{c}(N_{c}-2)D_{SSV}(m_{\omega},m_{\omega},0)
+ 2(N_{c} -1) D_{SSV}(m_{\tilde{t}_{L1}},m_{\tilde{t}_{L2}},m_{G})\nonumber \\
 &+& {N_{c}-1\over N_{c}} D_{SSV}(m_{\tilde{t}_{L1}},m_{\tilde{t}_{L1}},\overline{m}_{G})
+ 
{1\over N_{c}} D_{SSV}(m_{\tilde{t}_{L2}},m_{\tilde{t}_{L2}},\overline{m}_{G})\nonumber \\
&+&
N_{c}(N_{c}-2) D_{SSV}(m_{\tilde{t}_{L2}},m_{\tilde{t}_{L2}},0)\nonumber \\
&+& 2(N_{c} -1) D_{SSV}(m_{\tilde{b}_{L1}},m_{\tilde{b}_{L2}},m_{G})
 + {N_{c}-1\over N_{c}} D_{SSV}(m_{\tilde{b}_{L1}},m_{\tilde{b}_{L1}},\overline{m}_{G})\nonumber \\
& +& 
{1\over N_{c}} D_{SSV}(m_{\tilde{b}_{L2}},m_{\tilde{b}_{L2}},\overline{m}_{G})+
N_{c}(N_{c}-2) D_{SSV}(m_{\tilde{b}_{L2}},m_{\tilde{b}_{L2}},0)\nonumber \\
&+& 2(N_{c} -1) D_{SSV}(m_{\tilde{b}_{R1}},m_{\tilde{b}_{R2}},m_{G})
 + {N_{c}-1\over N_{c}} D_{SSV}(m_{\tilde{b}_{R1}},m_{\tilde{b}_{R1}},\overline{m}_{G})\nonumber \\
& +& {1\over N_{c}} D_{SSV}(m_{\tilde{b}_{R2}},m_{\tilde{b}_{R2}},\overline{m}_{G})+
N_{c}(N_{c}-2) D_{SSV}(m_{\tilde{b}_{R2}},m_{\tilde{b}_{R2}},0)],
\label{SSVchi}
\end{eqnarray}

\begin{eqnarray}
(VVS) &=& -g_{s}^{2}{m_{G}^{2} \over 8}[(N_{c} -1) D_{VVS}(m_{G},m_{G},m_{u})
+ 2 {(N_{c}-1)^{2}\over N^{2}} D_{VVS}(\overline{m}_{G},\overline{m}_{G},m_{u})\nonumber \\
&+& N_{c}(N_{c}-2) D_{VVS}(0,m_{G},m_{\omega}) + {(N_{c}-2)^{2}\over N_{c}}
D_{VVS}(m_{G},\overline{m}_{G},m_{\omega})],
\label{VVSchi}
\end{eqnarray}

\begin{eqnarray}
(SV)& =& -{g_{s}^{2}\over 8}[2N_{c}(N_{c}-2) D_{SV}(m_{\omega},0) +
(N_{c} -1)[3D_{SV}(m_{\omega},m_{G}) + D_{SV}(m_{u},m_{G})] \nonumber \\
&+& {1\over N_{c}}[(N_{c} + 1)D_{SV}(m_{\omega},\overline{m}_{G})
+(N_{c}-1)D_{SV}(m_{u},\overline{m}_{G})] \nonumber \\
&+& 2N_{c}(N_{c}-2) D_{SV}(m_{\tilde{t}_{L2}},0) +
(N_{c} -1)[2D_{SV}(m_{\tilde{t}_{L2}},m_{G}) +2 D_{SV}(m_{\tilde{t}_{L1}},m_{G})] \nonumber \\
&+& {1\over N_{c}}[2D_{SV}(m_{\tilde{t}_{L2}},\overline{m}_{G})
+2D_{SV}(m_{\tilde{t}_{L1}},\overline{m}_{G})] \nonumber \\
&+& 2N_{c}(N_{c}-2) D_{SV}(m_{\tilde{b}_{L2}},0) +
(N_{c} -1)[2D_{SV}(m_{\tilde{b}_{L2}},m_{G}) +2 D_{SV}(m_{\tilde{b}_{L1}},m_{G})] \nonumber \\
&+& {1\over N_{c}}[2D_{SV}(m_{\tilde{b}_{L2}},\overline{m}_{G})
+2D_{SV}(m_{\tilde{b}_{L1}},\overline{m}_{G})] \nonumber \\
&+& 2N_{c}(N_{c}-2) D_{SV}(m_{\tilde{b}_{R2}},0) +
(N_{c} -1)[2D_{SV}(m_{\tilde{b}_{R2}},m_{G}) +2 D_{SV}(m_{\tilde{b}_{R1}},m_{G})] \nonumber \\
&+& {1\over N_{c}}[2D_{SV}(m_{\tilde{b}_{R2}},\overline{m}_{G})
+2D_{SV}(m_{\tilde{b}_{R1}},\overline{m}_{G})] \nonumber \\
&-& {3\over 8} g_{w}^{2}[D_{SV}(m_{\tilde{t}_{L1}},0) + D_{SV}(m_{\tilde{t}_{L2}},0) + D_{SV}(m_{\tilde{t}_{L3}},0) \nonumber \\
&+& D_{SV}(m_{\tilde{b}_{L1}},0) + D_{SV}(m_{\tilde{b}_{L2}},0) + D_{SV}(m_{\tilde{b}_{L3}},0)],
\label{SVchi}
\end{eqnarray}

\begin{eqnarray}
(SSS) &=& -\lambda_{U}^{2} \chi^{2}[3D_{SSS}(m_{u},m_{u},m_{u}) +
(2N_{c}-1)D_{SSS}(m_{u},m_{\omega},m_{\omega})] \nonumber \\
&-&\chi^{2}[ {1\over 4} h_{t}^{2}\sin^{2}\beta[D_{SSS}(m_{u},m_{h},m_{h}) +
3D_{SSS}(m_{u}, m_{\pi},m_{\pi})] \nonumber \\
&-& {\chi^{2}\over 2}[(h_{t}^{2} -{1\over 3} g_{s}^{2})^{2}
D_{SSS}(m_{u},m_{\tilde{t}_{L1}},m_{\tilde{t}_{L1}}) \nonumber \\
&+&
2(h_{t}^{2} -{1\over 2} g_{s}^{2})^{2}D_{SSS}(m_{\omega},m_{\tilde{t}_{L1}},m_{\tilde{t}_{L2}})
+ 2({g_{s}^{2}\over 6})^{2}D_{SSS}(m_{u},m_{\tilde{t}_{L2}},m_{\tilde{t}_{L2}}) \nonumber \\
&+&(h_{t}^{2} -{1\over 3} g_{s}^{2})^{2}
D_{SSS}(m_{u},m_{\tilde{b}_{L1}},m_{\tilde{b}_{L1}})\nonumber \\
&+&
2(h_{t}^{2} -{1\over 2} g_{s}^{2})^{2}D_{SSS}(m_{\omega},m_{\tilde{b}_{L1}},m_{\tilde{b}_{L2}})
+ 2({g_{s}^{2}\over 6})^{2}D_{SSS}(m_{u},m_{\tilde{b}_{L2}},m_{\tilde{b}_{L2}}) \nonumber \\
&+&({1\over 3} g_{s}^{2})^{2}
D_{SSS}(m_{u},m_{\tilde{b}_{R1}},m_{\tilde{b}_{R1}}) +
2({1\over 2} g_{s}^{2})^{2}D_{SSS}(m_{\omega},m_{\tilde{b}_{R1}},m_{\tilde{b}_{R2}})\nonumber \\
&+& 2({g_{s}^{2}\over 6})^{2}D_{SSS}(m_{u},m_{\tilde{b}_{R2}},m_{\tilde{b}_{R2}})],
\label{SSSchi}
\end{eqnarray}

\begin{eqnarray}
(SS) &=& {1\over 4}\biggl({g_{w}^{2}\over 8} + {g_{s}^{2}\over 6}\biggr)[8 D_{SS}(m_{\tilde{t}_{L1}},m_{\tilde{t}_{L1}}) +
24 D_{SS}(m_{\tilde{t}_{L2}},m_{\tilde{t}_{L2}}) + 16 D_{SS}(m_{\tilde{t}_{L1}},m_{\tilde{t}_{L2}})\nonumber \\
&+& 8 D_{SS}(m_{\tilde{b}_{L1}},m_{\tilde{b}_{L1}}) +
24 D_{SS}(m_{\tilde{b}_{L2}},m_{\tilde{b}_{L2}}) + 16 D_{SS}(m_{\tilde{b}_{L1}},m_{\tilde{b}_{L2}})] \nonumber \\
&+& {1\over 4}( {g_{s}^{2}\over 6})[8 D_{SS}(m_{\tilde{t}_{R1}},m_{\tilde{t}_{R1}}) +
24 D_{SS}(m_{\tilde{t}_{R2}},m_{\tilde{t}_{R2}}) + 16 D_{SS}(m_{\tilde{t}_{R1}},m_{\tilde{t}_{R2}})\nonumber \\
&+& 8 D_{SS}(m_{\tilde{b}_{R1}},m_{\tilde{b}_{R1}}) +
24 D_{SS}(m_{\tilde{b}_{R2}},m_{\tilde{b}_{R2}}) + 16 D_{SS}(m_{\tilde{b}_{R1}},m_{\tilde{b}_{R2}})]\nonumber \\
&+& h_{t}^{2}[D_{SS}(m_{\tilde{t}_{L1}},m_{\tilde{t}_{R1}}) + D_{SS}(m_{\tilde{t}_{L2}},m_{\tilde{t}_{R2}})
+ D_{SS}(m_{\tilde{t}_{L3}},m_{\tilde{t}_{R3}})] \nonumber \\
& +& h_{t}^{2}[D_{SS}(m_{\tilde{b}_{L1}},m_{\tilde{t}_{R1}}) + D_{SS}(m_{\tilde{b}_{L2}},m_{\tilde{t}_{R2}})
+ D_{SS}(m_{\tilde{b}_{L3}},m_{\tilde{t}_{R3}})]\nonumber \\
&+& \biggl({1\over 2} h_{t}^{2} + {1\over 8} g_{w}^{2}\cos 2\beta\biggr)[D_{SS}(m_{\tilde{t}_{L1}},m_{h})
+ D_{SS}(m_{\tilde{t}_{L2}},m_{h})  + D_{SS}(m_{\tilde{t}_{L3}},m_{h}) \nonumber \\
&+& 2(D_{SS}(m_{\tilde{b}_{L1}},m_{\pi})
+ D_{SS}(m_{\tilde{b}_{L2}},m_{\pi})  + D_{SS}(m_{\tilde{b}_{L3}},m_{\pi}))\nonumber \\
&+&D_{SS}(m_{\tilde{t}_{L1}},m_{\pi})
+ D_{SS}(m_{\tilde{t}_{L2}},m_{\pi})  + D_{SS}(m_{\tilde{t}_{L3}},m_{\pi}) ]\nonumber \\
&- &  \biggl( {1\over 8} g_{w}^{2}\cos 2\beta\biggr)[D_{SS}(m_{\tilde{b}_{L1}},m_{h})
+ D_{SS}(m_{\tilde{b}_{L2}},m_{h})  + D_{SS}(m_{\tilde{b}_{L3}},m_{h}) \nonumber \\
&+& 2(D_{SS}(m_{\tilde{t}_{L1}},m_{\pi})
+ D_{SS}(m_{\tilde{t}_{L2}},m_{\pi})  + D_{SS}(m_{\tilde{t}_{L3}},m_{\pi}))\nonumber \\
&+&D_{SS}(m_{\tilde{b}_{L1}},m_{\pi})
+ D_{SS}(m_{\tilde{b}_{L2}},m_{\pi})  + D_{SS}(m_{\tilde{b}_{L3}},m_{\pi}) ]\nonumber \\
&+& {1\over 2} h_{t}^{2}\sin^{2}\beta[D_{SS}(m_{\tilde{t}_{R1}},m_{h})
+ D_{SS}(m_{\tilde{t}_{R2}},m_{h})  + D_{SS}(m_{\tilde{t}_{R3}},m_{h}) \nonumber \\
&+& 3(D_{SS}(m_{\tilde{t}_{R1}},m_{\pi})
+ D_{SS}(m_{\tilde{t}_{R2}},m_{\pi})  + D_{SS}(m_{\tilde{t}_{R3}},m_{\pi}))]
+ {g_{w}^{2}\over 4} (2-N_{c})[D_{SS}(m_{\tilde{t}_{L1}},m_{\tilde{b}_{L1}})\nonumber \\
& +& D_{SS}(m_{\tilde{t}_{L2}},m_{\tilde{b}_{L2}}) + D_{SS}(m_{\tilde{t}_{L3}},m_{\tilde{b}_{L3}})].
\label{SSchi}
\end{eqnarray}
The counterterm contribution to the mass term comes from an analogous
contribution to that of eq. (\ref{countermphi}) and is\footnote{
There is an additional counterterm contribution in the $\chi$-direction
for the case of a light higgsino.}

\begin{equation}
\delta V = {T^{2}\over 16 \pi^{2}} {\chi^{2}\over 2}{146\over 27}g_{s}^{4}.
\end{equation}

\subsection{Integration over the heavy scale}

The second part of the calculation arises, as noticed in the paper by Kajantie et al. \cite{KLRS}: when  the ``heavy" particles have been integrated out
 their contributions
to the 3D mass parameters  should also be included, as they can substantially
vary the value of the parameters $\Lambda_{H_{3}}, \Lambda_{U_{3}}$. In order to do this we must
calculate the 2-loop contributions to the effective potential in the $\phi$-
and $\chi$-directions from the heavy fields: $Q, D, C_{o}, A_{o}$

For the $\phi$-direction the expression of the effective potential
to 2-loops from the heavy particles can be deduced from the expressions
in the paper
by Espinosa \cite{Espinosa}.
For the $\chi$-direction this is new.
The masses in the shifted theory are now given by

\begin{equation}
m_{\tilde{t}_{L_{1}}}^{2} = \overline{m}_{Q_{3}}^{2} + (h_{t}^{L} + \Lambda_{3} + 
\Lambda_{4}^{s}){\phi^{2}\over 2} + (h_{t}^{QU} + g_{s_{1}}^{QU} + g_{s_{2}}^{QU})
{\chi^{2}\over 2},
\label{mmtL1}
\end{equation}

\begin{equation}
m_{\tilde{t}_{L_{2,3}}}^{2} = \overline{m}_{Q_{3}}^{2} + (h_{t}^{L} + \Lambda_{3} + 
\Lambda_{4}^{s}){\phi^{2}\over 2} + (g_{s_{2}}^{QU})
{\chi^{2}\over 2},
\label{mmtL2}
\end{equation}

\begin{equation}
m_{\tilde{b}_{L_{1}}}^{2} = \overline{m}_{Q_{3}}^{2} + ( \Lambda_{3} + 
\Lambda_{4}^{c}){\phi^{2}\over 2} + (h_{t}^{QU} + g_{s_{1}}^{QU} + g_{s_{2}}^{QU})
{\chi^{2}\over 2},
\label{mmbL1}
\end{equation}

\begin{equation}
m_{\tilde{b}_{L_{2,3}}}^{2} = \overline{m}_{Q_{3}}^{2} + (\Lambda_{3} + 
\Lambda_{4}^{c}){\phi^{2}\over 2} + (g_{s_{2}}^{QU})
{\chi^{2}\over 2},
\label{mmbL2}
\end{equation}

\begin{equation}
m_{\tilde{b}_{R_{1}}}^{2} = m_{D_{3}}^{2} + (g_{s_{1}}^{UD} + g_{s_{2}}^{UD})
{\chi^{2}\over 2},
\label{mmbR1}
\end{equation}

\begin{equation}
m_{\tilde{b}_{R_{2,3}}}^{2} = \overline{m}_{D_{3}}^{2}  + (g_{s_{2}}^{QU})
{\chi^{2}\over 2}.
\label{mmbR2}
\end{equation}
The expressions for the rest of the fields are given in \cite{Laine2}.
The 2-loop contributions from the heavy scale are given below. We stress
that the $D$-integrals in eqs. (\ref{Vheavyphi}) and (\ref{V2heavy}) are just $3D$ integrals, our notation
follows that of refs. \cite{Laine2, farakos, KLRS}\footnote{Our convention
for the functions $ D_{VVV}, D_{VVS}$ is that of \cite{Laine2}.}.

\subsubsection{$\phi$-direction}

\begin{eqnarray}
V_{2}^{heavy} &=& {3\over 8} g_{w_{3}}^{2}[D_{LS}(m_{h},m_{A_{0}}) + 3 D_{LS}(m_{\pi},m_{A_{0}})] 
+ 3 g_{w_{3}}^{2} D_{LV}(m_{W},m_{A_{0}})
\nonumber \\
&-&{3 \over 16} g_{w_{3}}^{2} \phi^{2} D_{LLS}(m_{h},m_{A_{0}},m_{A_{0}})
 -{3\over 2} g_{w_{3}}^{2}D_{LLV}(m_{W},m_{A_{0}},m_{A_{0}})\nonumber
\\ 
& -& {g_{w_{3}}^{2}\over 8} N_{c}[D_{SSV}(m_{\tilde{t}_{L}},m_{\tilde{t}_{L}},m_{W})
+ D_{SSV}(m_{\tilde{b}_{L}},m_{\tilde{b}_{L}},m_{W}) +
 4 D_{SSV}(m_{\tilde{t}_{L}},m_{\tilde{b}_{L}},m_{W})] \nonumber \\
&-&{g_{s_{3}}^{2}\over 4}(N_{c}^{2} -1)[D_{SSV}(m_{\tilde{t}_{L}},m_{\tilde{t}_{L}},0) +
D_{SSV}(m_{\tilde{b}_{L}},m_{\tilde{b}_{L}},0) 
+
D_{SSV}(m_{\tilde{b}_{R}},m_{\tilde{b}_{R}},0)]\nonumber \\
& -& \bigl[(h_{t}^{L}  + \Lambda_{3} + \Lambda_{4}^{s})^{2}
D_{SSS}(m_{\tilde{t}_{L}},m_{\tilde{t}_{L}},m_{h})
+ (\Lambda_{3} + \Lambda_{4}^{c})^{2} D_{SSS}(m_{\tilde{b}_{L}},m_{\tilde{b}_{L}},m_{h})
\nonumber \\
&+&  (h_{t}^{L}-\Lambda_{4}^{c} + \Lambda_4^{s})^{2}
D_{SSS}(m_{\tilde{t}_{L}},m_{\tilde{b}_{L}},m_{\pi})\bigr]{\phi^{2}\over 2}N_{c}\nonumber \\
& -& {1\over 4} g_{s_{3}}^{2}(N_{c}^{2} -1)[D_{SV}(m_{\tilde{t}_{L}},0) +
 D_{SV}(m_{\tilde{b}_{L}},0)  +
D_{SV}(m_{\tilde{b}_{R}},0)]\nonumber \\
&-& {3\over 8} g_{w_{3}}^{2}N_{c}[D_{SV}(m_{\tilde{t}_{L}},m_{W})+
D_{SV}(m_{\tilde{b}_{L}},m_{W})]\nonumber \\
 &+& (2\Lambda_{1} + g_{s_{1}}^{QQ} + g_{s_{2}}^{QQ}) N_{c}(2- N_{c}) D_{SS}(m_{\tilde{t}_{L}},m_{\tilde{b}_{L}})\nonumber \\
&+&
(h_{t}^{QU} + g_{s_{1}}^{QU} + 3 g_{s_{2}}^{QU})N_{c}[D_{SS}(m_{\tilde{t}_{L}},m_{\tilde{t}_{R}}) +
D_{SS}(m_{\tilde{b}_{L}},m_{\tilde{t}_{R}})] \nonumber \\
&+& (\Lambda_{1} + \lambda_{Q_{3}})N_{c}(N_{c}+
1)[D_{SS}(m_{\tilde{t}_{L}},m_{\tilde{t}_{L}}) +
D_{SS}(m_{\tilde{b}_{L}},m_{\tilde{b}_{L}})] \nonumber \\
&+& \lambda_{D_{3}} N_{c}(N_{c}+1)[ D_{SS}(m_{\tilde{b}_{R}},m_{\tilde{b}_{R}})] \nonumber \\
&+& N_{c}{1\over 2} (h_{t}^{L} + \Lambda_{3} + \Lambda_{4}^{s})[
D_{SS}(m_{\tilde{t}_{L}},m_{h}) + 2D_{SS}(m_{\tilde{b}_{L}},m_{\pi})
+ D_{SS}(m_{\tilde{t}_{L}},m_{\pi})] \nonumber \\
&+& {1\over 2} N_{c}(\Lambda_{3} + \Lambda_{4}^{c})[D_{SS}(m_{\tilde{b}_{L}},m_{h}) + 2
D_{SS}(m_{\tilde{t}_{L}},m_{\pi}) + D_{SS}(m_{\tilde{b}_{L}},m_{\pi})]\nonumber \\
&+& (g_{s_{1}}^{QD} + 3 g_{s_{2}}^{QD}) N_{c}[D_{SS}(m_{\tilde{t}_{L}},m_{\tilde{b}_{R}}) +
D_{SS}(m_{\tilde{b}_{L}}, m_{\tilde{b}_{R}})] \nonumber \\
& + &(g_{s_{1}}^{UD} + 3 g_{s_{2}}^{UD}) N_{c}[D_{SS}(m_{\tilde{t}_{R}},m_{\tilde{b}_{R}})].
\label{Vheavyphi}
\end{eqnarray}

\subsubsection{$\chi$-direction}

\begin{eqnarray}
V_{2}^{heavy} &=& {g_{s_{3}}^{2}\over 4}\biggl((N_{c}-1) [D_{LS}(m_{u},m_{C_{0}}) + D_{LS}(m_{\omega},m_{C_{0}})]\nonumber \\
&+& {1\over N_{c}}[4D_{LS}(m_{\omega},m_{C_{0}}) + 2D_{LS}(m_{u},m_{C_{0}})]\biggr) \nonumber \\
&-& g_{s_{3}}^{2}  {N_{c}\over 2}\biggl[(N_{c}-1) D_{LLS}(m_{C_{0}},m_{C_{0}},m_{u}) +
2{(N_{c}-1)^{2}\over N_{c}^{2}}D_{LLS}(m_{C_{0}},m_{C_{0}},m_{u}) \nonumber \\
&+&
N_{c}(N_{c}-2)D_{LLS}(0,m_{C_{0}},m_{\omega}) + {(N_{c}-2)^{2}\over
N_{c}}D_{LLS}(m_{C_{0}},m_{C_{0}},m_{\omega})\biggr] \nonumber \\
&-& g_{s_{3}}^{2} {N_{c}\over 2}[-D_{LV}(m_{C_{0}},\overline{m}_{G}) -D_{LV}(m_{C_{0}},m_{G})
- (N_{c}-1)D_{LV}(m_{C_{0}},m_{G})] \nonumber \\
&-&g_{s_{3}}^{2}{N_{c}\over 4}[D_{LLV}(m_{C_{0}},m_{C_{0}},\overline{m}_{G}) + 2
D_{LLV}(m_{C_{0}},m_{C_{0}},m_{G})\nonumber \\
& +& D_{LLV}(m_{C_{0}},m_{C_{0}},0)
+2D_{LLV}(m_{C_{0}},0,m_{G})]\nonumber \\
& -&{g_{w_{3}}^{2}\over 8}[D_{SSV}(m_{\tilde{t}_{L1}},m_{\tilde{t}_{L1}},0) +
D_{SSV}(m_{\tilde{t}_{L2}},m_{\tilde{t}_{L2}},0) +
D_{SSV}(m_{\tilde{t}_{L3}},m_{\tilde{t}_{L3}},0)\nonumber \\
& +&
D_{SSV}(m_{\tilde{b}_{L1}},m_{\tilde{b}_{L1}},0) +
D_{SSV}(m_{\tilde{b}_{L2}},m_{\tilde{b}_{L2}},0) +
D_{SSV}(m_{\tilde{b}_{L3}},m_{\tilde{b}_{L3}},0)\nonumber \\
& +&
4(D_{SSV}(m_{\tilde{t}_{L1}},m_{\tilde{b}_{L1}},0) + D_{SSV}(m_{\tilde{t}_{L2}},m_{\tilde{b}_{L2}},0) +
D_{SSV}(m_{\tilde{t}_{L3}},m_{\tilde{b}_{L3}},0) )]\nonumber \\
 &-& g_{s_{3}}^{2} {1\over 4}\biggl[
+ 2(N_{c} -1) D_{SSV}(m_{\tilde{t}_{L1}},m_{\tilde{t}_{L2}},m_{G})
 + {N_{c}-1\over N_{c}} D_{SSV}(m_{\tilde{t}_{L1}},m_{\tilde{t}_{L1}},\overline{m}_{G})\nonumber \\
&+& 
{1\over N_{c}} D_{SSV}(m_{\tilde{t}_{L2}},m_{\tilde{t}_{L2}},\overline{m}_{G})+
N_{c}(N_{c}-2) D_{SSV}(m_{\tilde{t}_{L2}},m_{\tilde{t}_{L2}},0)\nonumber \\
&+& 2(N_{c} -1) D_{SSV}(m_{\tilde{b}_{L1}},m_{\tilde{b}_{L2}},m_{G})
 + {N_{c}-1\over N_{c}} D_{SSV}(m_{\tilde{b}_{L1}},m_{\tilde{b}_{L1}},\overline{m}_{G})\nonumber \\
& +& 
{1\over N_{c}} D_{SSV}(m_{\tilde{b}_{L2}},m_{\tilde{b}_{L2}},\overline{m}_{G})\nonumber \\
&+&
N_{c}(N_{c}-2) D_{SSV}(m_{\tilde{b}_{L2}},m_{\tilde{b}_{L2}},0)
+ 2(N_{c} -1) D_{SSV}(m_{\tilde{b}_{R1}},m_{\tilde{b}_{R2}},m_{G})\nonumber \\
& +& {N_{c}-1\over N_{c}} D_{SSV}(m_{\tilde{b}_{R1}},m_{\tilde{b}_{R1}},\overline{m}_{G})\nonumber \\
& +& {1\over N_{c}} D_{SSV}(m_{\tilde{b}_{R2}},m_{\tilde{b}_{R2}},\overline{m}_{G})+
N_{c}(N_{c}-2) D_{SSV}(m_{\tilde{b}_{R2}},m_{\tilde{b}_{R2}},0)\biggr] \nonumber \\
&-& {\chi^{2}\over 2}[(h_{t}^{QU} + g_{s_{1}}^{QU} + g_{s_{2}}^{QU})^{2}
D_{SSS}(m_{u},m_{\tilde{t}_{L1}},m_{\tilde{t}_{L1}}) +
2(h_{t}^{QU} + g_{s_{1}}^{QU})^{2}D_{SSS}(m_{\omega},m_{\tilde{t}_{L1}},m_{\tilde{t}_{L2}})
\nonumber \\
&+& 2(g_{s_{2}}^{QU})^{2}D_{SSS}(m_{u},m_{\tilde{t}_{L2}},m_{\tilde{t}_{L2}})
+ (h_{t}^{QU} + g_{s_{1}}^{QU} + g_{s_{2}}^{QU})^{2}
D_{SSS}(m_{u},m_{\tilde{b}_{L1}},m_{\tilde{b}_{L1}}) \nonumber \\
& +& 2(h_{t}^{QU} + g_{s_{1}}^{QU})^{2}D_{SSS}(m_{\omega},m_{\tilde{b}_{L1}},m_{\tilde{b}_{L2}})
+ 2(g_{s_{2}}^{QU})^{2}D_{SSS}(m_{u},m_{\tilde{b}_{L2}},m_{\tilde{b}_{L2}}) \nonumber \\
&+&(  g_{s_{1}}^{UD} + g_{s_{2}}^{UD})^{2}
D_{SSS}(m_{u},m_{\tilde{b}_{R1}},m_{\tilde{b}_{R1}}) +
2(g_{s_{1}}^{UD})^{2}D_{SSS}(m_{\omega},m_{\tilde{b}_{R1}},m_{\tilde{b}_{R2}})\nonumber \\
&+& 2(g_{s_{2}}^{UD})^{2}D_{SSS}(m_{u},m_{\tilde{b}_{R2}},m_{\tilde{b}_{R2}})] \nonumber \\
&-&{g_{s_{3}}^{2}\over 8}\biggl[2N_{c}(N_{c}-2) D_{SV}(m_{\tilde{t}_{L2}},0) +
(N_{c} -1)[2D_{SV}(m_{\tilde{t}_{L2}},m_{G}) +2 D_{SV}(m_{\tilde{t}_{L1}},m_{G})] \nonumber \\
&+& {1\over N_{c}}[2D_{SV}(m_{\tilde{t}_{L2}},\overline{m}_{G})
+2D_{SV}(m_{\tilde{t}_{L1}},\overline{m}_{G})] \nonumber \\
&+& 2N_{c}(N_{c}-2) D_{SV}(m_{\tilde{b}_{L2}},0) +
(N_{c} -1)[2D_{SV}(m_{\tilde{b}_{L2}},m_{G}) +2 D_{SV}(m_{\tilde{b}_{L1}},m_{G})] \nonumber \\
&+& {1\over N_{c}}[2D_{SV}(m_{\tilde{b}_{L2}},\overline{m}_{G})
+2D_{SV}(m_{\tilde{b}_{L1}},\overline{m}_{G})] \nonumber \\
&+& 2N_{c}(N_{c}-2) D_{SV}(m_{\tilde{b}_{R2}},0) +
(N_{c} -1)[2D_{SV}(m_{\tilde{b}_{R2}},m_{G}) +2 D_{SV}(m_{\tilde{b}_{R1}},m_{G})] \nonumber \\
&+& {1\over N_{c}}[2D_{SV}(m_{\tilde{b}_{R2}},\overline{m}_{G})
+2D_{SV}(m_{\tilde{b}_{R1}},\overline{m}_{G})]\biggr] \nonumber \\
&-& {3\over 8} g_{w_{3}}^{2}[D_{SV}(m_{\tilde{t}_{L1}},0) + D_{SV}(m_{\tilde{t}_{L2}},0) + D_{SV}(m_{\tilde{t}_{L3}},0)\nonumber \\
& +& D_{SV}(m_{\tilde{b}_{L1}},0) + D_{SV}(m_{\tilde{b}_{L2}},0) + D_{SV}(m_{\tilde{b}_{L3}},0)]
\nonumber \\
 &+& {1\over 4}(\Lambda_{1} + \lambda_{Q_{3}})[8 D_{SS}(m_{\tilde{t}_{L1}},m_{\tilde{t}_{L1}}) +
24 D_{SS}(m_{\tilde{t}_{L2}},m_{\tilde{t}_{L2}}) + 16 D_{SS}(m_{\tilde{t}_{L1}},m_{\tilde{t}_{L2}})\nonumber \\
&+& 8 D_{SS}(m_{\tilde{b}_{L1}},m_{\tilde{b}_{L1}}) +
24 D_{SS}(m_{\tilde{b}_{L2}},m_{\tilde{b}_{L2}}) + 16 D_{SS}(m_{\tilde{b}_{L1}},m_{\tilde{b}_{L2}})]\nonumber \\
&+& {1\over 4}(\lambda_{D_{3}})[ 8 D_{SS}(m_{\tilde{b}_{R1}},m_{\tilde{b}_{R1}}) +
24 D_{SS}(m_{\tilde{b}_{R2}},m_{\tilde{b}_{R2}}) + 16 D_{SS}(m_{\tilde{b}_{R1}},m_{\tilde{b}_{R2}})]\nonumber \\
&+& (h_{t}^{QU} + g_{s_{1}}^{QU})[D_{SS}(m_{\tilde{t}_{L1}},m_{\tilde{t}_{R1}}) + D_{SS}(m_{\tilde{t}_{L2}},m_{\tilde{t}_{R2}})
+ D_{SS}(m_{\tilde{t}_{L3}},m_{\tilde{t}_{R3}})]\nonumber \\
&+& ( h_{t}^{QU} + g_{s_{1}}^{QU})[D_{SS}(m_{\tilde{b}_{L1}},m_{\tilde{t}_{R1}}) + D_{SS}(m_{\tilde{b}_{L2}},m_{\tilde{t}_{R2}})
+ D_{SS}(m_{\tilde{b}_{L3}},m_{\tilde{t}_{R3}})]\nonumber \\
&+& g_{s_{2}}^{QU}[D_{SS}(m_{\tilde{t}_{L1}},m_{\tilde{t}_{R1}})+
D_{SS}(m_{\tilde{t}_{L1}},m_{\tilde{t}_{R2}}) + D_{SS}(m_{\tilde{t}_{L1}},m_{\tilde{t}_{R3}})\nonumber \\
&+& D_{SS}(m_{\tilde{t}_{L2}},m_{\tilde{t}_{R1}}) + D_{SS}(m_{\tilde{t}_{L2}},m_{\tilde{t}_{R2}})+
D_{SS}(m_{\tilde{t}_{L2}},m_{\tilde{t}_{R3}})\nonumber \\
& +& D_{SS}(m_{\tilde{t}_{L3}},m_{\tilde{t}_{R1}})
+ D_{SS}(m_{\tilde{t}_{L3}},m_{\tilde{t}_{R2}})
+ D_{SS}(m_{\tilde{t}_{L3}},m_{\tilde{t}_{R3}}) \nonumber \\&+&
D_{SS}(m_{\tilde{b}_{L1}},m_{\tilde{t}_{R1}})+
D_{SS}(m_{\tilde{b}_{L1}},m_{\tilde{t}_{R2}}) + D_{SS}(m_{\tilde{b}_{L1}},m_{\tilde{t}_{R3}})\nonumber \\
&+& D_{SS}(m_{\tilde{b}_{L2}},m_{\tilde{t}_{R1}}) + D_{SS}(m_{\tilde{b}_{L2}},m_{\tilde{t}_{R2}})+
D_{SS}(m_{\tilde{b}_{L2}},m_{\tilde{t}_{R3}}) + D_{SS}(m_{\tilde{b}_{L3}},m_{\tilde{t}_{R1}})\nonumber \\
&+& D_{SS}(m_{\tilde{b}_{L3}},m_{\tilde{t}_{R2}})
+ D_{SS}(m_{\tilde{b}_{L3}},m_{\tilde{t}_{R3}})]\nonumber \\
&+& {1\over 2}( h_{t}^{L} + \Lambda_{3} + \Lambda_{4}^{s})[D_{SS}(m_{\tilde{t}_{L1}},m_{h})
+ D_{SS}(m_{\tilde{t}_{L2}},m_{h})  + D_{SS}(m_{\tilde{t}_{L3}},m_{h}) \nonumber \\
&+& 2(D_{SS}(m_{\tilde{b}_{L1}},m_{\pi})
+ D_{SS}(m_{\tilde{b}_{L2}},m_{\pi})  + D_{SS}(m_{\tilde{b}_{L3}},m_{\pi}))\nonumber \\
&+&D_{SS}(m_{\tilde{t}_{L1}},m_{\pi})
+ D_{SS}(m_{\tilde{t}_{L2}},m_{\pi})  + D_{SS}(m_{\tilde{t}_{L3}},m_{\pi}) ]\nonumber \\
&+ &  ( \Lambda_{3} + \Lambda_{4}^{c})[D_{SS}(m_{\tilde{b}_{L1}},m_{h})
+ D_{SS}(m_{\tilde{b}_{L2}},m_{h})  + D_{SS}(m_{\tilde{b}_{L3}},m_{h}) \nonumber \\
&+& 2(D_{SS}(m_{\tilde{t}_{L1}},m_{\pi})
+ D_{SS}(m_{\tilde{t}_{L2}},m_{\pi})  + D_{SS}(m_{\tilde{t}_{L3}},m_{\pi}))\nonumber \\
&+& D_{SS}(m_{\tilde{b}_{L1}},m_{\pi})
+ D_{SS}(m_{\tilde{b}_{L2}},m_{\pi})  + D_{SS}(m_{\tilde{b}_{L3}},m_{\pi}) ]\nonumber \\
&+& 2\Lambda_{1} (2-N_{c})[D_{SS}(m_{\tilde{t}_{L1}},m_{\tilde{b}_{L1}}) +
D_{SS}(m_{\tilde{t}_{L2}},m_{\tilde{b}_{L2}}) + D_{SS}(m_{\tilde{t}_{L3}},m_{\tilde{b}_{L3}})]\nonumber \\
&+&  g_{s_{1}}^{QD}[D_{SS}(m_{\tilde{t}_{L1}},m_{\tilde{b}_{R1}}) + D_{SS}(m_{\tilde{t}_{L2}},m_{\tilde{b}_{R2}})
+ D_{SS}(m_{\tilde{t}_{L3}},m_{\tilde{b}_{R3}})]\nonumber \\
&+&  g_{s_{1}}^{QD}[D_{SS}(m_{\tilde{b}_{L1}},m_{\tilde{b}_{R1}}) + D_{SS}(m_{\tilde{b}_{L2}},m_{\tilde{b}_{R2}})
+ D_{SS}(m_{\tilde{b}_{L3}},m_{\tilde{b}_{R3}})]\nonumber \\
&+& g_{s_{2}}^{QD}[D_{SS}(m_{\tilde{t}_{L1}},m_{\tilde{b}_{R1}})+
D_{SS}(m_{\tilde{t}_{L1}},m_{\tilde{b}_{R2}}) + D_{SS}(m_{\tilde{t}_{L1}},m_{\tilde{b}_{R3}})\nonumber \\
&+& D_{SS}(m_{\tilde{t}_{L2}},m_{\tilde{b}_{R1}}) + D_{SS}(m_{\tilde{t}_{L2}},m_{\tilde{b}_{R2}})+
D_{SS}(m_{\tilde{t}_{L2}},m_{\tilde{b}_{R3}})\nonumber \\
& +& D_{SS}(m_{\tilde{t}_{L3}},m_{\tilde{b}_{R1}})
+ D_{SS}(m_{\tilde{t}_{L3}},m_{\tilde{b}_{R2}})\nonumber \\
&+& D_{SS}(m_{\tilde{t}_{L3}},m_{\tilde{b}_{R3}}) +
D_{SS}(m_{\tilde{b}_{L1}},m_{\tilde{b}_{R1}})+
D_{SS}(m_{\tilde{b}_{L1}},m_{\tilde{b}_{R2}}) + D_{SS}(m_{\tilde{b}_{L1}},m_{\tilde{b}_{R3}})\nonumber \\
&+& D_{SS}(m_{\tilde{b}_{L2}},m_{\tilde{b}_{R1}}) + D_{SS}(m_{\tilde{b}_{L2}},m_{\tilde{b}_{R2}})+
D_{SS}(m_{\tilde{b}_{L2}},m_{\tilde{b}_{R3}}) + D_{SS}(m_{\tilde{b}_{L3}},m_{\tilde{b}_{R1}})\nonumber \\
&+& D_{SS}(m_{\tilde{b}_{L3}},m_{\tilde{b}_{R2}})
+ D_{SS}(m_{\tilde{b}_{L3}},m_{\tilde{b}_{R3}})]\nonumber \\
&+&  g_{s_{1}}^{UD}[D_{SS}(m_{\tilde{t}_{R1}},m_{\tilde{b}_{R1}}) + D_{SS}(m_{\tilde{t}_{R2}},m_{\tilde{b}_{R2}})
+ D_{SS}(m_{\tilde{t}_{R3}},m_{\tilde{b}_{R3}})]\nonumber \\
&+& g_{s_{2}}^{UD}[D_{SS}(m_{\tilde{t}_{R1}},m_{\tilde{b}_{R1}})+
D_{SS}(m_{\tilde{t}_{R1}},m_{\tilde{b}_{R2}}) + D_{SS}(m_{\tilde{t}_{R1}},m_{\tilde{b}_{R3}})\nonumber \\
&+& D_{SS}(m_{\tilde{t}_{R2}},m_{\tilde{b}_{R1}}) + D_{SS}(m_{\tilde{t}_{R2}},m_{\tilde{b}_{R2}})+
D_{SS}(m_{\tilde{t}_{R2}},m_{\tilde{b}_{R3}})\nonumber \\
& +& D_{SS}(m_{\tilde{t}_{R3}},m_{\tilde{b}_{R1}})
+ D_{SS}(m_{\tilde{t}_{R3}},m_{\tilde{b}_{R2}})\nonumber \\
&+& D_{SS}(m_{\tilde{t}_{R3}},m_{\tilde{b}_{R3}})].
\label{V2heavy}
\end{eqnarray}

\subsubsection{Mass terms}

Using the results presented in the previous sections, we can finally write the full
expressions for the mass terms of eqs. (\ref{mH3mu2}) and (\ref{mU3mu2})

\begin{eqnarray}
\overline{m}_{U_{3}}^{2}(\mu) &=& m_{U}^{2}\biggl(1 + 4 g_{s}^{2} {L_{b}\over 16\pi^{2}}\biggr)
+ T \biggl({1\over 3} g_{s_{3}}^{2}
 + {2\over 3} \lambda_{U_{3}} + {1\over 6} \gamma_{3} +
{1\over 6}(h_{t}^{QU} + g_{s_{1}}^{QU} + 3 g_{s_{2}}^{QU} + g_{s_{1}}^{UD} + 3
g_{s_{2}}^{UD})\biggr)
\nonumber \\
&-& {L_{b}\over 16 \pi^{2}}\biggl({4\over 3} g_{s}^{2} m_{U}^{2} + 2 h_{t}^{2} \sin^{2}\beta( m_{H}^{2}+
m_{Q}^{2})\biggr)\nonumber \\
&+& {1\over (16 \pi^{2})}\biggl(8 \overline{g}_{s_{3}}^{4} + {64 \over 3} \overline{\lambda}_{U_{3}} \overline{g}_{s_{3}}^{2} - 16 \overline{\lambda}_{U_{3}}^{2} - 2 \overline{\gamma}_{3}^{2} + 3 \overline{g}_{w_{3}}^{2} \overline{\gamma}_{3}\biggr) 
\biggl( \log\biggl({3T\over \mu}\biggr) +c\biggr)\nonumber \\
& +& {T^{2}\over (16\pi^{2})}\biggl(g_{s}^{4}\biggl({146\over 27} + {2\over 3} + {11\over 54}\biggr)\biggr)
\nonumber \\
&+& {1\over (16 \pi^{2})}\biggl(-{1\over 4} g_{s_{3}}^{4}\biggl({29\over 9}
\biggl( \log\biggl({3T\over (2 m_{C_{o}})}\biggr) +c + {1\over 2}\biggr)
+ 3 \biggl( \log\biggl({3T\over m_{C_{o}}}\biggr) +c + {1\over 2}\biggr)\biggr)
\nonumber \\
&+& {21\over 4} g_{s_{3}}^{4}\biggl( \log\biggl({3T\over (2 m_{C_{o}})}\biggr) +c+ {1\over 2}\biggr)\biggr) \nonumber \\
&+& {1 \over (16 \pi^{2})} \biggl(( 2(h_{t}^{QU} + g_{s_{1}}^{QU} +
g_{s_{2}}^{QU})^{2} + 8(h_{t}^{QU} + g_{s_{1}}^{QU})^{2} + 4 (g_{s_{2}}^{QU})^{2})\biggl( \log\biggl({3T\over (2 m_{Q_{3}})}\biggr) +c + {1\over 2}\biggr)
\nonumber \\
&+& ((g_{s_{1}}^{UD} + g_{s_{2}}^{UD})^{2} + (g_{s_{2}}^{UD})^{2} + 4 (g_{s_{1}}^{UD})^{2})\biggl( \log\biggl({3T\over (2 m_{D_{3}})}\biggr) +c
 + {1\over 2}\biggr)
\nonumber \\
&+& {1\over 4} g_{s_{3}}^{2}\biggl( (8(g_{s_{3}}^{2} - 2(h_{t}^{QU} +
g_{s_{1}}^{QU} + g_{s_{2}}^{QU}) - 2 g_{s_{2}}^{QU}) + {4\over 3}( {2\over 3}
g_{s_{3}}^{2} - 4( h_{t}^{QU} + g_{s_{1}}^{QU} + g_{s_{2}}^{QU})) 
\nonumber \\
&+& {2\over 3}({2\over 3} g_{s_{3}}^{2} - 4 g_{s_{2}}^{QU}) - 24 g_{s_{2}}^{QU}
\biggr),
\label{mU3mu}
\end{eqnarray}

\begin{eqnarray}
\overline{m}_{H_{3}}^{2}(\mu) &=& m_{H}^{2}\biggl(1 + {9\over 4} g_{w}^{2} {L_{b}\over 16\pi^{2}} - 3 h_{t}^{2} {L_{f}
\over 16\pi^{2}}\biggr) +T\biggl({1\over 2}\lambda_{H_{3}} + {3\over 16}
 g_{w_{3}}^{2} + {1\over 16} g'^{2}T
\nonumber \\
&+& {1\over 4} h_{t}^{f} + {1\over 4}
(h_{t}^{L} + 2\Lambda_{3}^{Q} + \Lambda_{4}^{c} + \Lambda_{4}^{s}
 + \gamma_{3}\biggr)\nonumber\\
&-&{L_{b}\over 16\pi^{2}}\biggl(6 \lambda m_{H}^{2}+ 3(m_{Q}^{2} + m_{U}^{2})h_{t}^{2}\sin^{2}\beta\biggr)\nonumber \\
&+& {1\over (16\pi^{2})}\biggl({51\over 16} \overline{g}_{w_{3}}^{4} + 9 \overline{
\lambda}_{H_{3}} \overline{g}_{w_{3}}^{2} - 12 \overline{\lambda}_{H_{3}}^{2} - 3 \overline{\gamma}_{3}^{2} + 8 \overline{g}_{s_{3}}^{2}
 \overline{\gamma}_{3}\biggr)\biggl( \log\biggl({3T\over \mu}\biggr) +c\biggr)
\nonumber \\
&+& {T^{2}\over (16\pi^{2})}\biggl(g_{w}^{4}({137\over 96} + {9\over 2} \log 2
+ {1\over 4}) + {3\over 4}\lambda g_{w}^{2}\biggr) \nonumber \\
&+& {1\over (16\pi^{2})}\biggl( {15\over 8} g_{w_{3}}^{4} 
\biggl( \log\biggl({3T\over (2 m_{A_{o}})}\biggr) +c\biggr) + {9\over 16} g_{w_{3}}^{4}\biggr) + {T^{2}\over (16 \pi^{2})} \biggl({2\over 3} g_{s}^{2} h_{t}^{2} 
\sin^{2}\beta\biggr) \nonumber \\
&+& {1\over (16\pi^{2})}\biggl(-{3\over 8}g_{w_{3}}^{2}(3 g_{w_{3}}^{2} -12 (h_{t}^{L} + \Lambda_{3}^{Q} + \Lambda_{4}^{s}) - 12(\Lambda_{3}^{Q} +
 \Lambda_{4}^{c})) \biggl( \log\biggl({3T\over (2 m_{Q_{3}})}\biggr) +c + {1\over 2}
\biggr)\nonumber \\
&-& 2 g_{s_{3}}^{2}(-4(h_{t}^{L} + \Lambda_{3}^{c} + \Lambda_{4}^{s})
- 4(\Lambda_{3}^{Q} + \Lambda_{4}^{c}))\biggl( \log\biggl({3T\over (2 m_{Q_{3}})}\biggr) +c + {1\over 2}\biggr) \nonumber \\
&-& 3(( h_{t}^{L} + \Lambda_{3}^{Q} + \Lambda_{4}^{s})^{2} + 
(\Lambda_{3}^{Q} + \Lambda_{4}^{c})^{2} + (h_{t}^{L} - \Lambda_{4}^{c} 
+ \Lambda_{4}^{s})^{2})\biggl( \log\biggl({3T\over m_{Q_{3}}}\biggr)+c +
{1\over 2}\biggr)\biggr) \nonumber \\
&+& {T^{2}\over (16 \pi^{2})}\biggl({2\over 3} g_{s}^{2} h_{t}^{2} \sin^{2}\beta + {3\over 8} g_{w}^{2} h_{t}^{2} \sin^{2}\beta + {3\over 8} g_{w}^{4}
- {1\over 16} g_{w}^{4}\biggr),
\label{mH3mu}
\end{eqnarray}
where

\begin{eqnarray}
h_{t}^{f} &=& h_{t}^{2}\sin^{2}\beta T\biggl(1 - {3\over 8} {1\over (16 \pi^{2})}
\biggl[ \biggl( 12 h_{t}^{2} \sin^{2}\beta - 6 g_{w}^{2} - {64\over 3} g_{s}^{2}\biggr)L_{f} \nonumber \\
&+& g_{w}^{2}(2 + 28 \log 2) - 96 \lambda \log 2 + 16 h_{t}^{2} \sin^{2}\beta \log 2 - {64\over 9}
g_{s}^{2}(4\log 2 -3)\biggr]\biggr),\\
c &=& {1\over 2}\biggl[\ln {8\pi\over 9} + {\zeta'(2)\over \zeta(2)} - 2\gamma\biggr] \approx -0.348725.
\end{eqnarray}
The exact values of the parameters $\Lambda_{H_{3}}$ and $\Lambda_{U_{3}}$ 
in eqs. (\ref{mH3mu2}) and (\ref{mU3mu2}), depend on the particle content
of the theory and on the input parameters. In fact, the dependence on
$\tan\beta$ and $m_{t_{R}}$, for the range of values we are interested in,
is weak. The dependence on $m_{Q}$ and $m_{D}$ is stronger. For $m_{Q} = m_{D} = 300$ GeV, the corresponding values of the
parameters are in the range $\Lambda_{H_{3}} = (1.6-1.8) T$, and $\Lambda_{U_{3}} = (6.6-6.9)T$.
\subsection{Zero-Temperature Renormalization}

In order to complete the matching of the  3D parameters to the 4D
physical parameters, we must renormalize the zero-temperature theory.
We will not go into the details of the renormalization,
but refer the reader to the literature in which
the pole masses for the relevant particles of our calculation
have been obtained considering the full particle spectrum of the
MSSM \cite{bagger, donini}. We use the
expressions given in ref. \cite{bagger}, keeping only the
top Yukawa coupling, in the appropriate (large-$m_{A}$) limit. We have kept all
$\overline{\mu}$-dependent contributions to order $g_{i}^{4}(h_{t}^{4})$ but have neglected the constant contributions,
which are not multiplied by $h_{t}$ or $g_{s}$. In this way all explicit
dependence on  $\overline{\mu}$ is cancelled at one-loop when we relate
the 3D parameters to pole masses. 

Another zero-temperature constraint that we impose is the stability of the
physical vacuum. In principle, a metastable region exists in which the
colour-breaking minimum is lower than the physical one at zero temperature.
If the time for the transition to this lower minimum is greater than the age of
the Universe then this region of parameter space  is
also acceptable. However, we will not consider these issues in the
present paper.
 The constraint for absolute stability
 can be obtained by studying the effective potential
at zero temperature \cite{Carena1, Carena2}. This gives the constraint
$-m_{U}^{2} \leq  (m_{U}^{c})^{2}$, where

\begin{equation}
m_{U}^{c} = \biggl( {m_{h}^{2} v^{2} g_{s}^{2}\over 12}\biggr)^{1/4}.
\label{zeroTconst}
\end{equation}

\section{Results}

\begin{figure}[ht]
\vskip -10 pt
\epsfxsize=4in
\epsfysize=6in
\epsffile{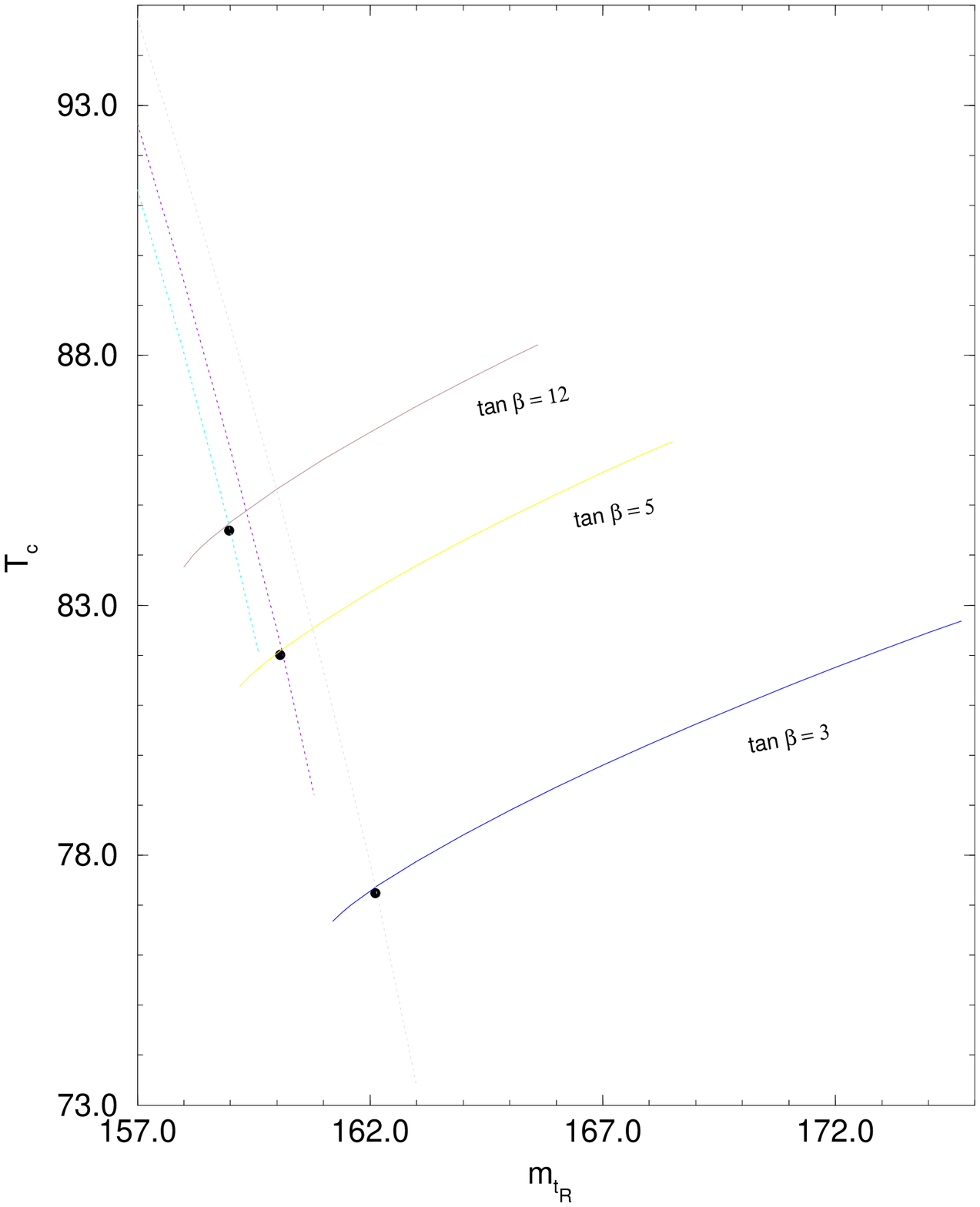}
\vskip -10 pt
\caption{Critical temperatures in the $\phi$- (solid) and $\chi$- (dotted) directions
as functions of $m_{\tilde{t}_{R}}$ for $\tan\beta =3,5,12$ and  $m_{Q} = 300$ GeV. }
\label{TcmtR}
\end{figure}

\begin{figure}
\vskip -10 pt
\epsfxsize=4in
\epsfysize=6in
\epsffile{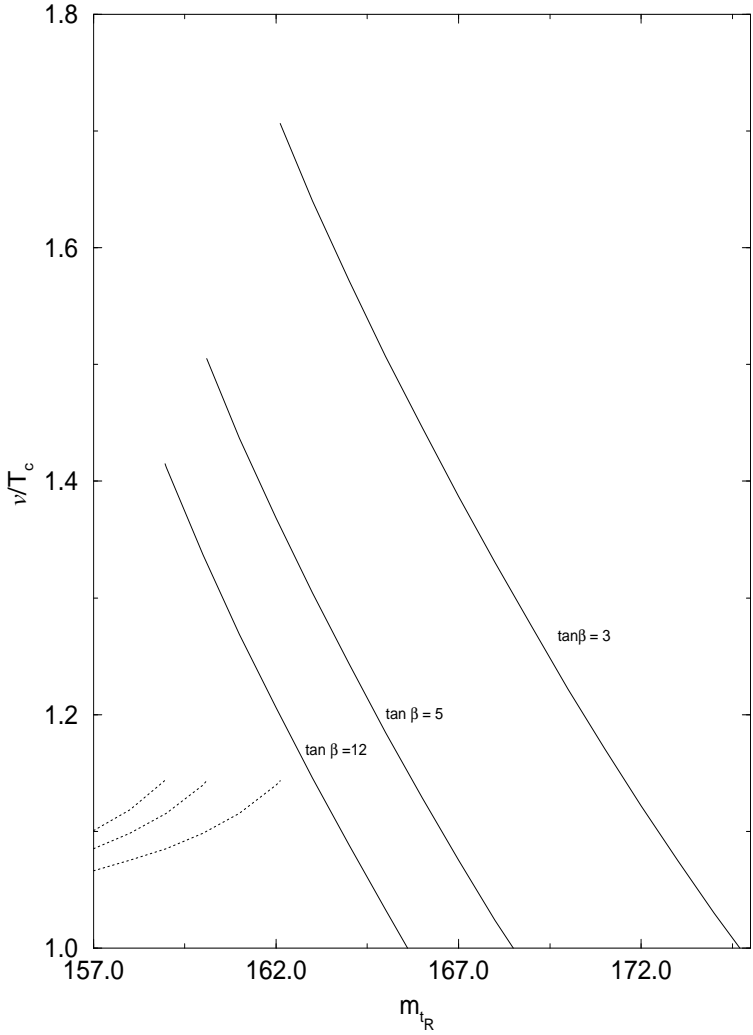}
\vskip -10 pt
\caption{Plot of ${v\over T}$ as a function of $m_{\tilde{t}_{R}}$ in the $\phi$- (solid line) and $\chi$- (dotted line) directions for $\tan\beta =3,5,12$ and $m_{Q} = 300$ GeV. For a given value of $\tan\beta$ the lines end at
the same value of the right handed stop mass.}
\label{vTmtR}
\end{figure}

With the previous results we can now analyse the phase transition.
 In fig. \ref{TcmtR} we
show the critical temperatures for the transitions in the $\phi$- and
$\chi$-directions as a function of the right handed stop pole mass $m_{\tilde{t}_{R}}$, for $\tan\beta =3,5,12$. We find that, for  $m_{Q} \sim 300$ GeV, there
 still is a region in which a two-stage phase transition can occur. This region
is to the left of the crossing points of the curves.
With respect to the work of ref. \cite{Laine2} the structure of the
phase diagram is preserved, although it is slightly shifted towards
higher
values of the right stop mass. Note that there is a considerable difference between our values of the critical temperatures and those in ref. \cite{Laine2}. Our analysis
concludes that the structure of the phase diagram is robust to small
additional corrections.
This structure  is maintained also
for $m_{Q} = m_{D} = 1$ TeV\footnote{In this case the third-generation left handed squark doublet and right handed sbottom are decoupled from the thermal bath. The 
relations obtained from the dimensional-reduction procedure can be
deduced from all of the formulae presented in the previous sections. All contributions
from these fields are suppressed at finite temperature; however, a residual
dependence on $m_{Q}$ as a consequence of a
zero-temperature effect persists for the scalar Higgs self coupling $\lambda$ \cite{Laine2, Laine3}.}.
The total effect  does not substantially increase or decrease 
 the range of values of the right handed stop mass for which a two-stage phase transition can occur. However, the exact location of this small range in the value
of $m_{\tilde{t}_{R}}$ depends on the value of the third-generation left handed squark doublet mass. In fig. \ref{vTmtR} we give
the values of ${v\over T}$ for three different values of $\tan\beta$. As expected, the strength of the phase transition
has a weak dependence on the values of the scales that have been fixed in our calculation, and only slight differences
are observed with respect to previous analyses.

\begin{figure}
\vskip -180 pt
\epsfxsize=4in
\epsfysize=6in
\epsffile{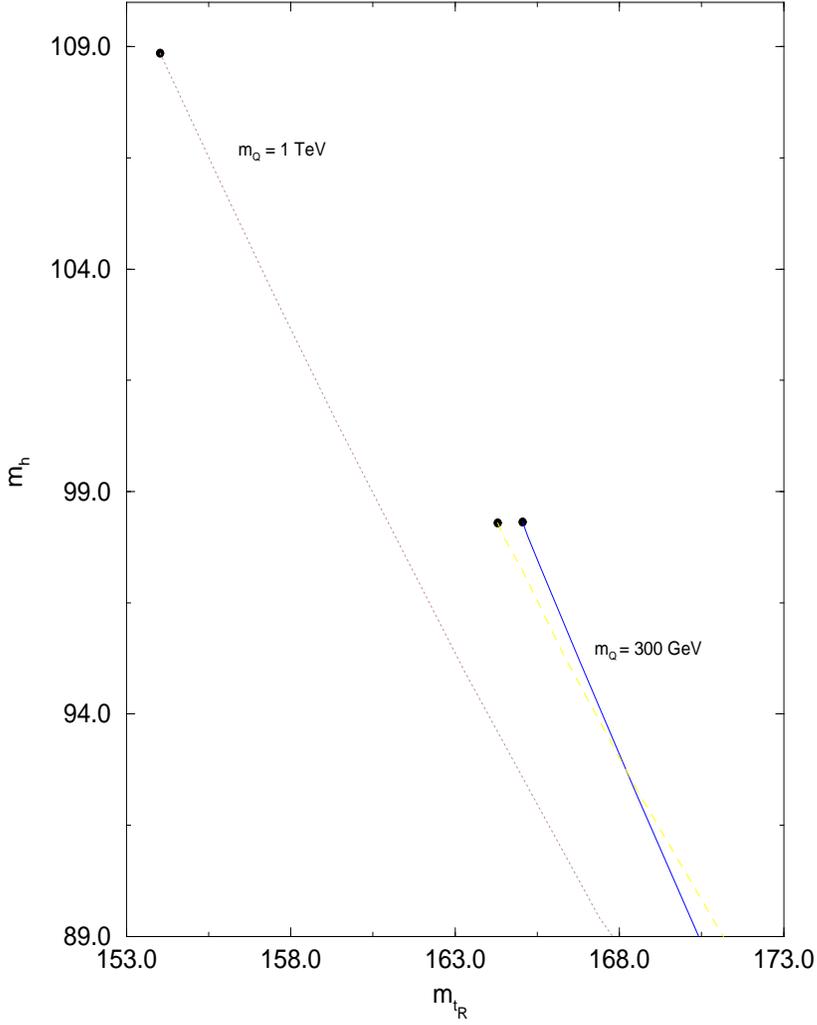}
\vskip -10 pt
\caption{Contours of ${v\over T} =1$ in the $m_{h}$-$m_{\tilde{t}_{R}}$ plane. The solid (dotted) line
corresponds to the results obtained within our approximations for $m_{Q}= 300$
GeV ($1$TeV).
The dashed line is the result using the approximations of ref. \cite{Laine2} for $m_{Q} = 300$ GeV. The region to the
left of the lines gives a sufficiently strong first-order phase transition, for a
given value of $m_{Q}$. }
\label{alpha}
\end{figure}

Figure \ref{alpha} shows lines of ${v\over T} =1$ for three different cases in the
$m_{h}$--$m_{\tilde{t}_{R}}$ plane. The phase transition is sufficiently strong
for  electroweak baryogenesis to the left of the solid (dotted) line
for $m_{Q}=300$ GeV ($m_{Q} = 1$ TeV), using the results obtained in this paper. The dashed line is the result
using the approximations of ref. \cite{Laine2}, for $m_{Q} = 300$ GeV. We can see that the full effect
of the corrections we have included on the strength of the phase transition is small.
The end-points of the lines correspond to the maximum value of the Higgs mass that is reached by the  effect of the zero-temperature radiative
corrections for a given value of $m_{Q}$, 
and the $y$-axis starts at the value of  the experimental limit on the Higgs mass\footnote{As we are working in the large-$m_{A}$ limit, the Standard Model bound on
 the Higgs mass is used.} \cite{moriond}.

The allowed region in parameter space is shown in figs. 4 and 5, given the
current experimental limits on the Higgs mass, for two
different values of $m_{Q}$. The region on the left of the solid line indicates
when a sufficiently strong first-order phase transition occurs. The
dotted line gives the condition for absolute stability of the
physical vacuum.  As explained above, to the left of this line the colour-breaking minimum is
lower than the physical one at zero-temperature. The dashed line is obtained
when the critical temperatures of the transitions in the $\phi$- and $\chi$-directions
are the same. A two-stage phase transition occurs to the left of the dashed line.
Note that, unlike the results of ref. \cite{Carena2} for $m_{Q}=1$ TeV and
zero
squark mixing, the
dashed and solid lines do not intersect.

As mentioned in the introduction, for a sufficiently heavy right handed stop field,
an effective theory with a single light scalar Higgs doublet field can be constructed.
In the appendix we give the formulae that  modify the equations presented
in the previous sections. There are two ways of constructing this
effective theory. We can either integrate out the right handed stop field
simultaneously with the other heavy fields, or perform a third stage of reduction
and integrate out the right stop field separately. The basic change in
the equations of the appendix will be to replace the 3D coupling and masses by the 
3D barred couplings and masses of section 2.2. It is clear, as mentioned above,
that the results obtained with these effective potentials are unreliable 
for low values of the right handed stop mass, as perturbation theory is no
longer under control. In terms of the 3D parameters we note in particular
that the 
difference between $\lambda_{H_{3}}$ and $\overline{\lambda}_{H_{3}}$ is
considerable, owing to the difference in $m_{U_{3}}^{2}$ and $\overline{m}_{U_{3}}^{2}$.
However, comparing the results obtained with these three separate 
approximations, we can obtain a more precise value of the right handed stop mass,
for which the effective
theory with a single light scalar field is a valid description.
Recall that the one-loop estimate gave as a lower limit  the value $m_{\tilde{t}_{R}} \gsi 177$ GeV \cite{Laine, Losada1}.
In fig. 6 we present the plot of the ratio of the vacuum expectation
value of the Higgs doublet to the temperature as a function of the 
right handed stop mass obtained for the three
approximations for the effective potential mentioned above. In this plot we have taken $\tan\beta=5$. The solid line
is the result obtained when using the 2-loop effective potential derived in the main
part of the paper. The dotted line corresponds to the results obtained 
after integrating out the right stop field simultaneously with 
the other heavy fields. The dashed line is the result obtained when a separate third stage of
reduction is performed. We see that only for very large values of the
right handed stop mass are the results  basically the same. As we move towards
lower values of $m_{t_{R}}$, the lines very quickly  start to diverge.
In particular, we see that for $m_{\tilde{t}_{R}}=177$ GeV ($m_{U} \sim 50$ GeV) 
the results using the effective potentials for a single light scalar
are completely unreliable. To understand why 
the results of the  dashed and dotted lines are so different
for smaller values of $m_{\tilde{t}_{R}}$, we recall that the strength of the transition
is dominated by the value of $\lambda_{H_{3}}$ in the final 3D effective theory for a single light scalar field at the phase transition.
The value of $\lambda_{H_{3}}$ is much smaller when
a third stage of reduction is performed, as the value of $m_{U_{3}}^{2}$ is much
smaller too. That is, the net effect artificially strengthens the first-order phase
transition.

\section{Conclusions}
We have performed a full two-loop dimensional reduction
of 4D MSSM parameters to the 3D couplings and masses of the effective theory.
In this way, we have fixed the scales appearing in the 3D mass terms
 that are due to the thermal polarizations and the
super-renormalizability of the 3D theory. The values of the
parameters $\Lambda_{H_{3}}$ and $\Lambda_{U_{3}}$ can vary 
significantly  for different values of the input parameters and
the particle content of the theory, thus modifying the critical temperatures of
the transitions.
We have compared our results with previous analyses. We conclude that the
corrections relevant to the preservation of the baryon asymmetry
are small and  that the allowed range of
masses is $m_{h} \lsi 110$ GeV and $m_{\tilde{t}_{R}} \lsi m_{t}$, in complete
agreement with previous results. We find that the phase structure diagram
still allows a possible two-stage phase transition for a small
range of values of $m_{\tilde{t}_{R}}$. This range of values is shifted
compared to previous results. 
However,  whether or not the transition actually occurs must be explicitly checked.
Initial lattice analysis suggests that the second stage of the transition
is extremely strong and thus  this transition might  not
have taken place on cosmological time scales. Consequently,  this region of parameter space
for electroweak baryogenesis would be excluded. From the comparison of the
results obtained using the different approximations to the
effective potential we can conclude that for $\tilde{m}_{t_{R}} \lsi 210$ GeV,
the  results obtained  for the strength of the phase transition with  a single light Higgs field at the transition point are unreliable.

\vspace{.3in}
\noindent
{\bf Acknowledgements}

\noindent
I would like to thank M. Shaposhnikov, C. Wagner
and in particular  M. Laine for many useful discussions. I especially thank
J.R. Espinosa for interesting discussions and comments on the
manuscript.

\appendix

\section*{Appendix}

\subsection*{Integrating out the right handed stop}
The additional corrections that  must  be included in the
case of integrating out the right handed stop are given below.
The Higgs self-coupling is modified by

\begin{equation}
\overline{\lambda}_{3} = \lambda_{3} - {3\over 16\pi} \gamma_{3}^{2} {1 \over
m_{U_{3}}}.
\label{newlambda}
\end{equation}
The one-loop correction to the mass term is

\begin{equation}
\overline{m}_{3}^{2} = m_{3}^{2} - {3\over 4\pi} \gamma_{3} m_{U_{3}}.
\label{newmH3}
\end{equation}

The relevant contributions from the 2-loop graphs have been given
in ref. \cite{Espinosa} with  the following modifications,
which include the effects of the dimensional-reduction procedure.
For the (SSV) and (SV) terms substitute $g_{s} \rightarrow g_{s_{3}}$,
for the (SSS) term substitute $h_{t}^{2}\sin^{2}\beta \rightarrow \gamma_{3}^{2}$ for the (SS) term substitute ${g_{s}^{2}\over 6} \rightarrow \lambda_{U_{3}}$ and  $h_{t}^{2}\sin^{2}\beta \rightarrow \gamma_{3}^{2}$.

\begin{figure}
\vskip -180 pt
\epsfxsize=4in
\epsfysize=6in
\epsffile{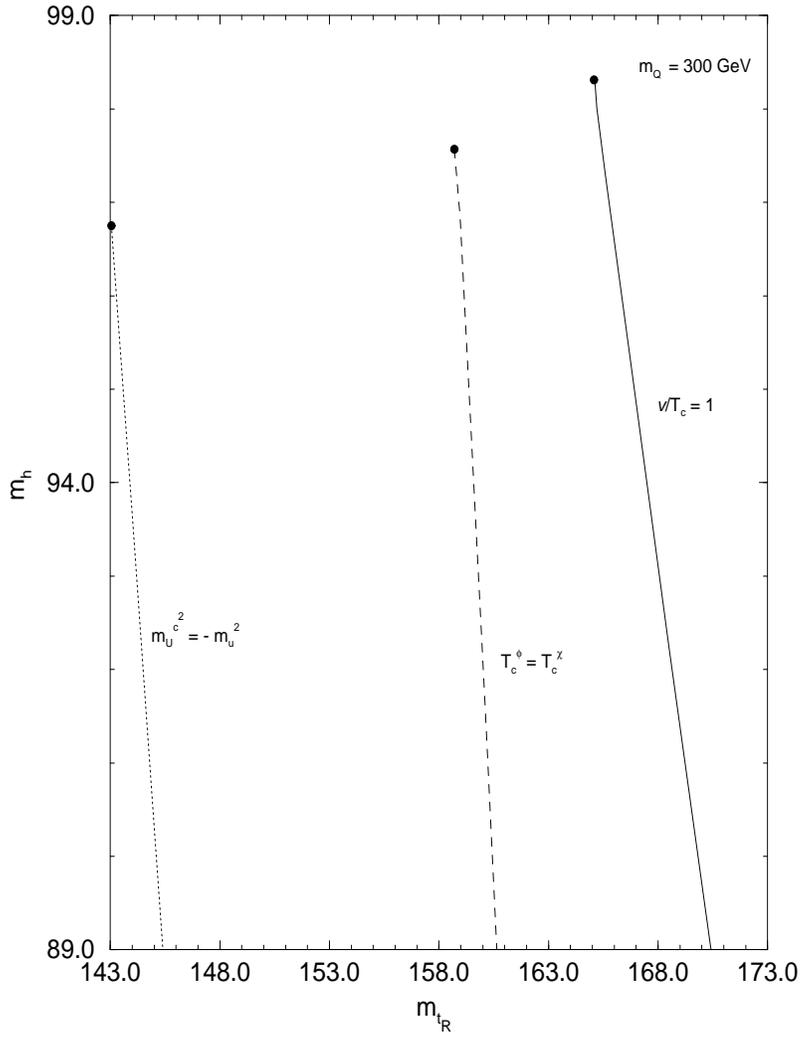}
\vskip -10 pt
\caption{Allowed region in $m_{h}$-$m_{\tilde{t}_{R}}$ plane for $m_{Q}=300$ GeV. To the left of the solid line
there is a sufficiently strong first-order phase transition, to the right of the dotted line
the physical vacuum is absolutely stable. The dashed line separates the region for which
a two-stage phase transition can occur. }
\label{comp1}
\end{figure}

\begin{figure}
\vskip -180 pt
\epsfxsize=4in
\epsfysize=6in
\epsffile{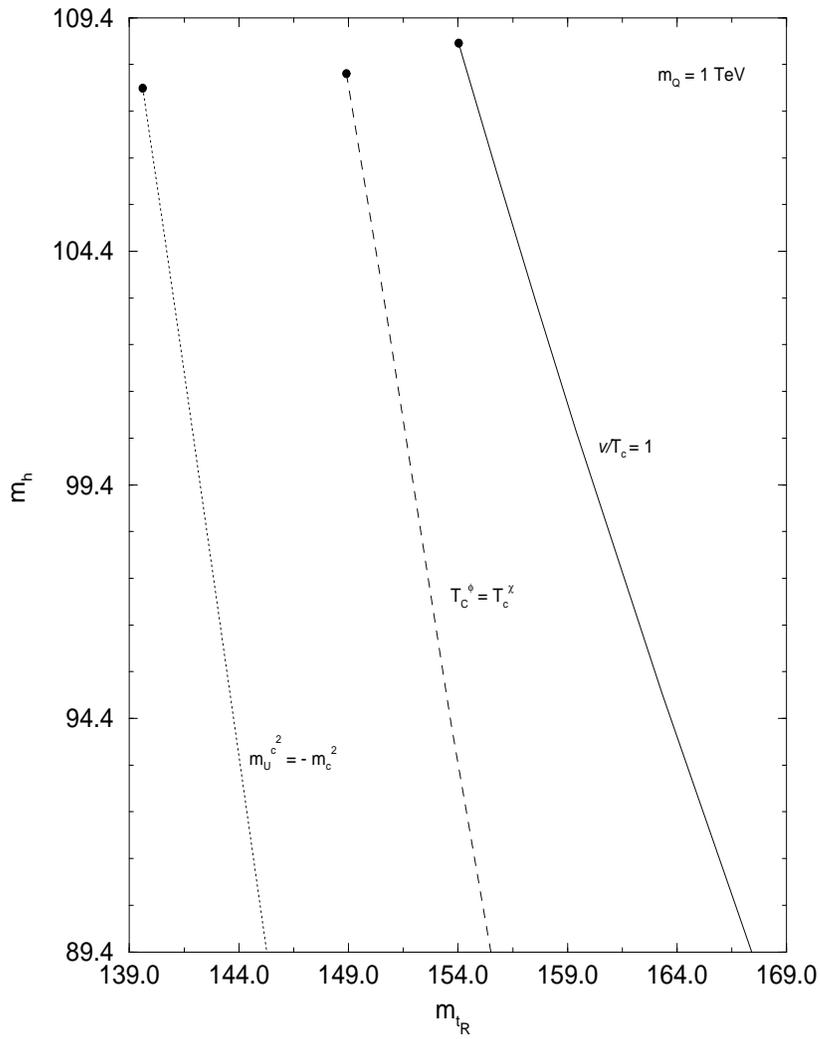}
\vskip -10 pt
\caption{Same as fig. \ref{comp1}, for $m_{Q}=1$TeV. }
\label{caren1}
\end{figure}

\begin{figure}
\vskip -180 pt
\epsfxsize=4in
\epsfysize=6in
\epsffile{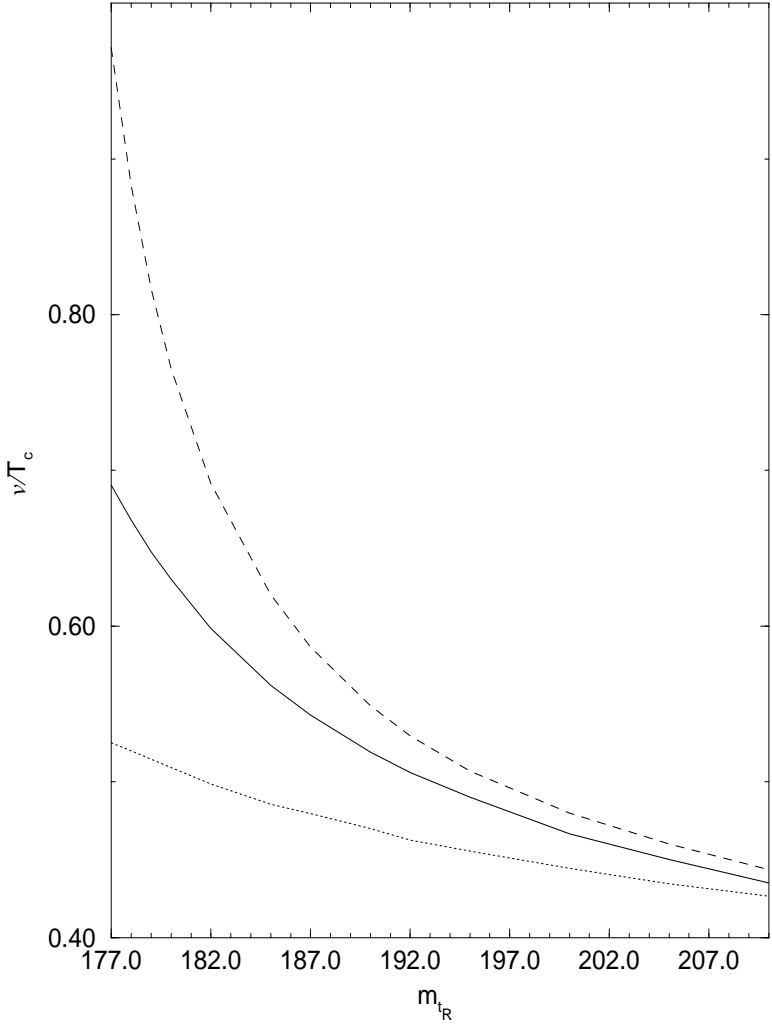}
\vskip -10 pt
\caption{Plot of ${v\over T}$ as a function of $m_{\tilde{t}_{R}}$ in the $\phi$-direction for three cases.
The solid line corresponds to the result obtained with the effective potential given in the main
part of the paper. The dotted line corresponds to the results obtained 
after integrating out the right stop field simultaneously with 
the other heavy fields. The dashed line is the result obtained when a separate third stage of
reduction is performed.
 }
\label{vteffec}
\end{figure}


\end{document}